\documentclass[iop,apj]{emulateapj}

\pdfoutput=1

\usepackage{apjfonts}
\usepackage[UKenglish]{isodate}
\usepackage{rotating}

\cleanlookdateon

\defcitealias{bellini2014}{Paper 1}
\defcitealias{harris1996}{H96}
\defcitealias{mclaughlin2005}{M05}

\newcommand {\wcen} {$\omega$~Centauri}
\newcommand {\kms} {km\,s$^{-1}$}
\newcommand {\masyr} {mas\,yr$^{-1}$}
\newcommand {\muasyr} {$\mu$as\,yr$^{-1}$}
\newcommand {\Msun} {M$_\odot$}

\def \Nused {N_\mathrm{used}}
\def \Nfound {N_\mathrm{found}}
\def \Nuf {\Nused / \Nfound}

\def \Qfit {\texttt{QFIT}}

\def \mux {\mu_\mathrm{x}}
\def \muy {\mu_\mathrm{y}}

\def \vi {v_\mathrm{i}}
\def \dvi {\delta v_\mathrm{i}}

\def \vtrial {\overline{v}_\mathrm{trial}}
\def \strial {\sigma_\mathrm{trial}}

\def \vbar {\overline{v}}
\def \vr {\overline{v}_\mathrm{r}}
\def \sr {\sigma_\mathrm{r}}
\def \vt {\overline{v}_\mathrm{t}}
\def \st {\sigma_\mathrm{t}}

\def \Nstar {N_\mathrm{star}}

\def \Rcore {R_\mathrm{core}}
\def \Rhalf {R_\mathrm{half}}

\def \score {\sigma_\mathrm{core}}
\def \shalf {\sigma_\mathrm{half}}

\def \smajor {\sigma_\mathrm{major}}
\def \sminor {\sigma_\mathrm{minor}}

\def \trelax {t_\mathrm{relax}}
\def \tbreak {t_\mathrm{break}}

\usepackage[breaklinks,colorlinks,citecolor=blue,linkcolor=blue,urlcolor=blue,pdftex,bookmarks=true]{hyperref}
\usepackage[all]{hypcap}

\hypersetup{
    pdftitle={HST PMs of GCs. II. Kinematic profiles and maps},
    pdfauthor={Laura L. Watkins, et al.}
}

\def\equationautorefname~#1\null{%
  equation~(#1)\null
}

\begin{document}

\title{\textit{Hubble Space Telescope} proper motion (HSTPROMO) catalogs of Galactic globular clusters$^{\ast}$. II. Kinematic profiles and maps}

\altaffiltext{$\ast$}{Based on proprietary and archival observations with the NASA/ESA Hubble Space Telescope, obtained at the Space Telescope Science Institute, which is operated by AURA, Inc., under NASA contract NAS 5-26555.}

\author{Laura~L.~Watkins, Roeland~P.~van~der~Marel, Andrea~Bellini, Jay~Anderson}

\affil{Space Telescope Science Institute, 3700 San Martin Drive, Baltimore MD 21218, USA}

\email{lwatkins@stsci.edu}

\slugcomment{Draft version, \today}
\shorttitle{HST proper motions in globular clusters. II. Kinematic profiles and maps}
\shortauthors{L.~L.~Watkins~et~al.}


\begin{abstract}
    We present kinematical analyses of 22 Galactic globular clusters using the \textit{Hubble Space Telescope} proper motion (HSTPROMO) catalogues recently presented in \citet{bellini2014}. For most clusters, this is the first proper-motion study ever performed, and, for many, this is the most detailed kinematic study of any kind. We use cleaned samples of bright stars to determine binned velocity-dispersion and velocity-anisotropy radial profiles and two-dimensional velocity-dispersion spatial maps. Using these profiles, we search for correlations between cluster kinematics and structural properties. We find that: (1) more centrally-concentrated clusters have steeper radial velocity-dispersion profiles; (2) on average, at 1$\sigma$ confidence in two dimensions, the photometric and kinematic centres of globular clusters agree to within $\sim 1"$, with a cluster-to-cluster rms of $4"$ (including observational uncertainties); (3) on average, the cores of globular clusters have isotropic velocity distributions to within 1\% ($\st/\sr = 0.992 \pm 0.005$), with a cluster-to-cluster rms of 2\% (including observational uncertainties); (4) clusters generally have mildly radially anisotropic velocity distributions ($\st/\sr \approx 0.8$--$1.0$) near the half-mass radius, with bigger deviations from isotropy for clusters with longer relaxation times; (5) there is a relation between $\sminor/\smajor$ and ellipticity, such that the more flattened clusters in the sample tend to be more anisotropic, with $\sminor/\smajor \approx 0.9$--$1.0$. Aside from these general results and correlations, the profiles and maps presented here can provide a basis for detailed dynamical modelling of individual globular clusters. Given the quality of the data, this is likely to provide new insights into a range of topics concerning globular cluster mass profiles, structure, and dynamics.
\end{abstract}



\section{Introduction}
\label{sect:introduction}

The internal kinematics of globular clusters hold the key to unlocking their mysteries. Different formation mechanisms and evolutionary paths leave unique imprints in their dynamical structures; using high-quality velocity data combined with detailed dynamical modelling, we can force clusters to yield up their secrets. Any such study hinges on the fact that stars trace the underlying potential in which they orbit and, so, are sensitive to mass both visible and invisible to our telescopes. From study of the motions of individual stars, we can piece together a coherent picture of the cluster as a whole.

Globular clusters are a common feature of massive galaxies, but they are also found associated with much smaller systems, such as dwarf galaxies. A number of clusters in the Milky Way \citep{bellazzini2003, martin2004, mackey2004, marinfranch2009} and nearby Andromeda \citep[M\,31,][]{mackey2010, mackey2013, huxor2014, veljanoski2014, mackey2014} are associated with tidal features, which suggests an extragalactic origin for those clusters, and it is likely that they formed in accreted dwarfs.

In fact, there is evidence that some globular clusters were not just formed in dwarf galaxies but are the stripped nuclei of dwarf galaxies that have been accreted by more massive hosts \citep[e.g.][]{freeman1993, meylan2001, bekki2003}. For example, globular cluster M\,54 (NGC\,6715) sits right at the heart of the Sagittarius (Sgr) dwarf spheroidal galaxy, though it shows clear kinematical differences from the coincident Sgr stars so it is believed to have formed in the halo of Sgr and sunk to the centre as a result of dynamical friction \citep{bellazzini2008, ibata2009}.

If this stripped-nucleus scenario is true, then theory predicts that globular clusters formed thus should still retain trace amounts of dark matter (although most will have been stripped away already), which would be readily detected in their kinematics. However, so far, dynamical studies suggest that globular-cluster dynamics can be entirely explained by accounting for the mass contained in stars \citep[e.g.][]{sollima2009, ibata2011}, and dark matter has yet to be definitively detected.

Some, though certainly not all, globular clusters are also suspected to harbour an intermediate-mass black hole (IMBH) at their centre, though results are conflicting. For example, \citet{noyola2008} and \citet{noyola2010} studied the centre of \wcen\ (NGC\,5139) using integral-field line-of-sight velocities and claimed to detect an IMBH of $\sim 4 \times 10^4$~\Msun; however \citet{vandermarel2010} performed a proper-motion study of the same region and ruled out such a massive black hole with high confidence. A more detailed study is ongoing (Watkins et al. in prep). Again, dynamical studies are key here as an IMBH will increase the velocity dispersion of nearby stars over what would be expected from the stars alone. However, it should be noted that mass segregation with no central IMBH will also increase the central velocity dispersion, as is likely the case for M15 \citep[e.g.][]{denbrok2014}. X-ray searches have so far been inconclusive; for example, \citet{haggard2013} failed to detect emission associated with an IMBH in \wcen, but this could imply either that there is no IMBH in \wcen\ or simply that accretion onto the IMBH is highly inefficient.

Stars in globular clusters undergo a series of two-body interactions; over time, the stars move towards a state of energy equipartition whereby all stars have the same kinetic energy. Recently, \citet{trenti2013} showed that clusters may never reach fully equipartition, but even partial equipartition will result in a system where the massive (bright) stars move more slowly than the less massive (faint) stars. Consequently, massive stars slowly sink towards the centre and less-massive stars tend to move outwards, a process known as mass segregation \citep{spitzer1969}. The resulting loss of kinetic energy from the core leads to gravothermal instability and, eventually, core collapse \citep[e.g.][]{lyndenbell1968}. A population of binaries (or even a single hard binary) near the cluster centre \citep{binney2008} or a central IMBH \citep{baumgardt2005} can halt core collapse as they act as an extra source of energy in the core.

Any kinematical analysis of a cluster will give us insight into its structure, formation and evolution, however proper motions have a number of advantages over line-of-sight velocities. Foremost, line-of-sight velocity measurements give us only one component of motion, so we are unable to determine velocity anisotropy. There is a well-known degeneracy between mass and velocity anisotropy, so dynamical studies using only line-of-sight velocities must make assumptions about the anisotropy in the system in order to make an estimate of the mass of the system. Proper motions, however, provide two components of motion, with which anisotropies on the plane of the sky can be determined. This provides critical information that can help to break the mass-anisotropy degeneracy and leads to more accurate results. Of course, the ideal case is to have both proper motions and line-of-sight velocities so that the full three-dimensional velocity distribution can be determined. Nevertheless, the information from proper motions alone offers an improvement over line-of-sight studies.

Another asset is that proper motions are measured by determining by how much stars have moved in images taken at different epochs, whereas line-of-sight velocities are determined from spectra, which are often possible to obtain for only the brightest stars. This difference in observing strategy means that proper-motion studies can typically go deeper than line-of-sight velocity observations; not only does this provide larger samples of stars, but also we can determine proper motions for stars of different mass. This is particularly important for studies of energy equipartition and mass segregation, and is a topic to which we will return in detail in future papers.

In \citet[][hereafter \citetalias{bellini2014}]{bellini2014}, we recently presented a set of \textit{Hubble Space Telescope} (\textit{HST}) proper-motion catalogues for 22 Milky Way globular clusters. These catalogues are the result of a search through archival \textit{HST} data to find fields in Galactic globular clusters that had been previously observed for other projects at multiple epochs, allowing us to measure proper motions. Thanks to both the stability and longevity of \textit{HST}, we were able to achieve exceptional precision over baselines of up to 12 years. The bright, well-measured stars have proper-motion measurements with median accuracy of 32~\muasyr and mean accuracy of 38~\muasyr. These values correspond to 1.2~\kms and 1.5~\kms at 8~kpc, a typical distance for the clusters in the sample.

Aside from the particular benefits of proper motions that we have already discussed, these catalogues have further value when considered as a population. With a sample of 22 clusters, we can look at statistics of the clusters as a whole and attempt to relate kinematic properties with photometric parameters.

Another advantage of our proper-motion catalogues is that the clusters are a very diverse set: they are found in different environments and are suspected to have very different structures and evolutionary histories. Some clusters are believed to be core-collapsed, while others are suspected to harbour intermediate-mass black holes whose presence would act to inhibit core-collapse (although, as previously discussed, the topic of IMBHs is still contentious and work is ongoing). NGC\,6715 (M\,54) lies right at the heart of the Sagittarius dwarf spheroidal and is clearly of extragalactic origin. Though with less direct evidence, some other clusters are also believed to have formed outside of our galaxy and been accreted, while others are thought to have been formed in situ. Some clusters in our sample are superimposed on dwarf galaxies in the Milky Way (namely, Sagittarius and the Small Magellanic Cloud), while others are relatively isolated. So there is a lot we can learn from both the similarities and the differences among the clusters.

Clearly, these catalogues have enormous potential for a wide range of very exciting science \citepalias[some of which we have barely touched on here, but see the introduction of][]{bellini2014}. We cannot hope to address all of the lingering questions surrounding globular clusters in one shot, but we can lay down a solid groundwork here, upon which we will build in future papers. We begin here with an analysis of the kinematical profiles and maps for each of the 22 clusters. 

\autoref{sect:catalogues} introduces the catalogues and discusses the cleaning that we perform in order to select high-quality samples of bright stars free of contaminants. In \autoref{sect:kinematics}, we briefly outline our kinematic estimation methods and, in \autoref{sect:results}, we derive one-dimensional and two-dimensional kinematic profiles. In \autoref{sect:discussion}, we provide context for our results via comparison with previous studies, and look for correlations with other cluster parameters and at statistics for the sample as a whole. We conclude in \autoref{sect:conclusions}.


\section{Cluster catalogues}
\label{sect:catalogues}

For our analysis, we use the \textit{HST} proper-motion catalogues for 22 Galactic globular clusters described in \citetalias{bellini2014}, to which we refer for detailed explanations of the data reduction and processing. Due to a lack of sufficient background sources to use as a frame of reference, the proper motions provided in the catalogues are relative to the bulk motion of the cluster. As explained in \citetalias{bellini2014}, we noted inhomogeneities in the mean velocity across each of the clusters, which we attribute to uncorrected charge-transfer-efficiency and errors in the geometric-distortion correction. To counteract these fluctuations, the catalogues also provide a set of local corrections, which can be applied to the proper motions to smooth out the mean velocity field. There is one exception: NGC\,7099, as the dataset contains insufficient stars with which to calculate local corrections. For the analysis in this paper, we apply these local corrections to all clusters for which they are available. So, by design, the mean velocity of cluster stars in a small area of the sky should be zero.

As we are only able to measure relative proper motions, the catalogues cannot be used to calculate bulk motions of the clusters through the Milky Way and they cannot be used to study internal solid-body rotation. Applying the local corrections will also remove any possible differential rotation. However, most importantly for the present study, the removal of any rotation signatures does not affect the dispersions that we measure. The data are therefore ideal for most dynamical applications we might wish to consider. But future dynamical models based on the proper motion data will have to make some assumptions about either the nature or absence of cluster rotation.\footnote{The removal of any rotation signatures will affect the azimuthal velocity second moment $\overline{v_\phi^2} = \sigma_\phi^2 + v_{\rm rot}^2$ that appears in the equations of hydrostatic equilibrium; this can only be estimated if we know or assume the rotation velocity $v_{\rm rot}$ independently. However, as most globular clusters are near circular, we expect that $v_{\rm rot} << \sigma_\phi$ in most cases (see \autoref{table:basics}). The only exception to this may be for highly flattened clusters, such as \wcen\ (NGC\,5139), which we know exhibits a high degree of rotation \citep{vandeven2006}. Furthermore, rotation is usually found in the outer regions of clusters while our data covers the inner regions, so we would expect any rotation signal to be low.}

\begin{table*}
    \caption{Characteristic quantities for our clusters.}
    \label{table:basics}
    \centering
    \begin{tabular}{cccccccccccc}
        \hline
        \hline
        Cluster & Other names & $\Rcore$ & $\Rhalf$ & $d$ & $\sigma_0$ & $\epsilon$ & $\log(t_\mathrm{core})$ & $\log(t_\mathrm{med})$  & $c$ & $\theta$ & $v_\mathrm{rot}$ \\
         & & (arcsec) & (arcsec) & (kpc) & (\kms) & & & & & (deg) & (\kms) \\
        \hline
        NGC\,104  & 47\,Tuc &  21.6 & 190.2 &   4.5 & 11.0 $\pm$ 0.3 &  0.09 &  7.84 &  9.55 & 2.01 $\pm$ 0.03 &    33 &  4.4 $\pm$ 0.4 \\
        NGC\,288  & \dots   &  81.0 & 133.8 &   8.9 &  2.9 $\pm$ 0.3 & \dots &  8.99 &  9.32 & 0.99 $\pm$ 0.04 & \dots &  0.5 $\pm$ 0.3 \\
        NGC\,362  & \dots   &  10.8 &  49.2 &   8.6 &  6.4 $\pm$ 0.3 &  0.01 &  7.76 &  8.93 & 1.80 $\pm$ 0.03 &    25 &      \dots     \\
        NGC\,1851 & \dots   &   5.4 &  30.6 &  12.1 & 10.4 $\pm$ 0.5 &  0.05 &  7.43 &  8.82 & 1.86 $\pm$ 0.04 &    86 &  1.6 $\pm$ 0.5 \\
        NGC\,2808 & \dots   &  15.0 &  48.0 &   9.6 & 13.4 $\pm$ 1.2 &  0.12 &  8.24 &  9.15 & 1.56 $\pm$ 0.03 &    31 &  3.3 $\pm$ 0.5 \\
        NGC\,5139 & \wcen   & 142.2 & 300.0 &   5.2 & 16.8 $\pm$ 0.3 &  0.17 &  9.60 & 10.09 & 1.31 $\pm$ 0.04 &    96 &  6.0 $\pm$ 1.0 \\
        NGC\,5904 & M\,5    &  26.4 & 106.2 &   7.5 &  5.5 $\pm$ 0.4 &  0.14 &  8.28 &  9.41 & 1.71 $\pm$ 0.03 &   124 &  2.6 $\pm$ 0.5 \\
        NGC\,5927 & \dots   &  25.2 &  66.0 &   7.7 &      \dots     &  0.04 &  8.39 &  8.94 &      \dots      &    63 &      \dots     \\
        NGC\,6266 & M\,62   &  13.2 &  55.2 &   6.8 & 14.3 $\pm$ 0.4 &  0.01 &  7.90 &  8.98 & 1.71 $\pm$ 0.03 &    52 &      \dots     \\
        NGC\,6341 & M\,92   &  15.6 &  61.2 &   8.3 &  6.0 $\pm$ 0.4 &  0.10 &  7.96 &  9.02 & 1.68 $\pm$ 0.03 &   141 &      \dots     \\
        NGC\,6362 & \dots   &  67.8 & 123.0 &   7.6 &  2.8 $\pm$ 0.4 &  0.07 &  8.80 &  9.20 & 1.09 $\pm$ 0.05 &    58 &      \dots     \\
        NGC\,6388 & \dots   &   7.2 &  31.2 &   9.9 & 18.9 $\pm$ 0.8 &  0.01 &  7.72 &  8.90 & 1.71 $\pm$ 0.03 &    56 &  3.9 $\pm$ 1.0 \\
        NGC\,6397 & \dots   &   3.0 & 174.0 &   2.3 &  4.5 $\pm$ 0.2 &  0.07 &  4.94 &  8.60 & 2.12 $\pm$ 0.04 &     2 &  0.2 $\pm$ 0.5 \\
        NGC\,6441 & \dots   &   7.8 &  34.2 &  11.6 & 18.0 $\pm$ 0.2 &  0.02 &  7.93 &  9.09 & 1.74 $\pm$ 0.03 &    74 & 12.9 $\pm$ 2.0 \\
        NGC\,6535 & \dots   &  21.6 &  51.0 &   6.8 &  2.4 $\pm$ 0.5 &  0.08 &  7.28 &  8.20 & 1.33 $\pm$ 0.16 &   105 &      \dots     \\
        NGC\,6624 & \dots   &   3.6 &  49.2 &   7.9 &  5.4 $\pm$ 0.5 &  0.06 &  6.61 &  8.71 &      \dots      &    89 &      \dots     \\
        NGC\,6656 & M\,22   &  79.8 & 201.6 &   3.2 &  7.8 $\pm$ 0.3 &  0.14 &  8.53 &  9.23 & 1.38 $\pm$ 0.04 &   139 &  1.5 $\pm$ 0.4 \\
        NGC\,6681 & M\,70   &   1.8 &  42.6 &   9.0 &  5.2 $\pm$ 0.5 &  0.01 &  5.82 &  8.65 & 2.01 $\pm$ 0.04 &   116 &      \dots     \\
        NGC\,6715 & M\,54   &   5.4 &  49.2 &  26.5 & 10.5 $\pm$ 0.3 &  0.06 &  8.24 &  9.93 & 2.04 $\pm$ 0.03 &    78 &  2.0 $\pm$ 0.5 \\
        NGC\,6752 & \dots   &  10.2 & 114.6 &   4.0 &  4.9 $\pm$ 0.4 &  0.04 &  6.88 &  8.87 & 2.07 $\pm$ 0.03 &   147 &  0.0 $\pm$ 0.0 \\
        NGC\,7078 & M\,15   &   8.4 &  60.0 &  10.4 & 13.5 $\pm$ 0.9 &  0.05 &  7.84 &  9.32 & 1.95 $\pm$ 0.03 &   125 &  3.8 $\pm$ 0.5 \\
        NGC\,7099 & M\,30   &   3.6 &  61.8 &   8.1 &  5.5 $\pm$ 0.4 &  0.01 &  6.37 &  8.88 & 2.12 $\pm$ 0.03 &    33 &  0.0 $\pm$ 0.0 \\
        \hline
    \end{tabular}
    
    \qquad
    
    \textbf{Notes.} Columns: (1) cluster identification in the NGC catalogue; (2) alternate names by which the cluster is known; (3) core radius; (4) half-light radius; (5) heliocentric distance; (6) central dispersion estimate; (7) ellipticity; (8) core relaxation time; (9) half-mass relaxation time; (10) King concentration; (11) position angle; (12) rotation amplitude.
    
    \textbf{References.} Columns: (3)-(9) \citetalias{harris1996}; (10) \citetalias{mclaughlin2005}; (11) \citet{white1987}; (12) \citet{bellazzini2012}.
\end{table*}

\autoref{table:basics} lists some basic statistics for our clusters, taken from \citet[][2010 edition, hereafter \citetalias{harris1996}]{harris1996}, \citet[][hereafter \citetalias{mclaughlin2005}]{mclaughlin2005} and \citet{white1987}. As well as giving a quick overview of the clusters for which we have proper-motion catalogues, we will use these quantities later to connect our kinematical analysis with previous photometric and kinematic studies.

In order to calculate accurate kinematics, it is vital that we have high-quality proper motions. Poorly-measured stars or those for which the uncertainties have been underestimated will bias our results, in particular, they will tend to increase the dispersions that we estimate. The datasets also contain contaminating stars from the Milky Way, and some contain further contamination from other nearby objects with which they are associated -- i.e. Sagittarius for NGC\,6624, NGC\,6681 and NGC\,6715, and the SMC for NGC\,104 and NGC\,362 -- which can further bias our results. To ensure that we have a reliable sample of proper motions and to remove outliers, we make a series of cuts on the datasets, as we now describe.

\begin{table*}
    \caption{Summary of catalogue cleaning.}
    \label{table:numbers}
    \centering
    \begin{tabular}{crrrrrccccc}
        \hline
        \hline
        Cluster & $N_\mathrm{catalogue}$ & $N_\mathrm{bright}$ & $N_\mathrm{quality1}$ & $N_\mathrm{quality2}$ & $N_\mathrm{final}$ & $M_\mathrm{cut}$ & $R_\mathrm{lim}$ & $R_\mathrm{in}$ & $R_\mathrm{out}$ & $\xi_\mathrm{cut}$ \\
        & & & & & & (mag) & (arcsec) & (arcsec) & (arcsec) &  \\
        (1) & (2) & (3) & (4) & (5) & (6) & (7) & (8) & (9) & (10) & (11) \\
        \hline
        NGC\,104  &  103638 &  28440 &  16199 &  16199 &  15999 & 18.5 & 40 &  0 & 20 & 100 \\
        NGC\,288  &   14970 &   2982 &   2891 &   2891 &   1267 & 19.9 & 40 &  0 & 20 & 100 \\
        NGC\,362  &   66766 &  16853 &  11237 &  10944 &  10469 & 19.8 & 40 &  0 & 20 &  98 \\
        NGC\,1851 &   63227 &  14320 &  11434 &   9217 &   7007 & 20.5 & 40 &  0 & 20 &  90 \\
        NGC\,2808 &   86420 &  38949 &  23957 &  17441 &  16829 & 20.6 & 40 &  0 & 20 &  80 \\
        NGC\,5139 &  313286 &  59446 &  46308 &  45795 &  45481 & 19.0 & 40 &  0 & 20 &  99 \\
        NGC\,5904 &   47627 &  10267 &   7293 &   6957 &   6886 & 19.4 & 40 &  0 & 20 &  95 \\
        NGC\,5927 &   60222 &  18153 &  15483 &  15483 &  13444 & 21.0 & 40 &  0 & 20 & 100 \\
        NGC\,6266 &   58272 &  28308 &  21400 &  17144 &  16558 & 20.0 & 40 &  0 & 20 &  88 \\
        NGC\,6341 &   83218 &  11416 &   8249 &   7884 &   7270 & 19.5 & 40 &  0 & 20 &  97 \\
        NGC\,6362 &    7951 &   3149 &   3007 &   3007 &   1387 & 19.7 & 40 &  0 & 20 & 100 \\
        NGC\,6388 &   86671 &  60365 &  41595 &  28504 &  11153 & 21.6 & 60 & 20 & 40 &  82 \\
        NGC\,6397 &   13593 &   2079 &   1932 &   1932 &   1812 & 17.4 & 40 &  0 & 20 & 100 \\
        NGC\,6441 &   84508 &  60993 &  32552 &  20774 &  10560 & 22.3 & 60 & 20 & 40 &  77 \\
        NGC\,6535 &    3348 &    885 &    821 &    821 &    147 & 20.1 & 40 &  0 & 30 & 100 \\
        NGC\,6624 &   13948 &   8442 &   6626 &   5686 &   1855 & 20.5 & 40 &  0 & 10 &  89 \\
        NGC\,6656 &   50622 &   8095 &   7232 &   7232 &   6984 & 18.5 & 40 &  0 & 20 & 100 \\
        NGC\,6681 &   24248 &   5641 &   5014 &   5014 &   4445 & 20.1 & 40 &  0 & 10 & 100 \\
        NGC\,6715 &   77190 &  49025 &  25494 &  21872 &   3887 & 22.6 & 40 &  0 & 10 &  94 \\
        NGC\,6752 &   38013 &   7684 &   5328 &   5328 &   5266 & 18.3 & 40 &  0 & 20 & 100 \\
        NGC\,7078 &   77837 &  18937 &  14479 &   8171 &   7822 & 20.2 & 60 & 10 & 30 &  69 \\
        NGC\,7099 &    2360 &    802 &    600 &    490 &    123 & 19.6 &  0 &  0 & 10 &  84 \\
        \hline
        Total     & 1377935 & 455231 & 309131 & 258786 & 196651 &      &    &    &    &     \\
        \hline
    \end{tabular}
    
    \qquad
    
    \textbf{Notes.} Columns: (1) cluster identification in the NGC catalogue; (2) total number of stars in catalogue; (3) number of stars remaining after magnitude cut; (4) number of stars remaining after quality cut on $\Nuf$ and $\chi^2$; (5) number of stars remaining after cut on $\Qfit$ or rms; (6) number of stars remaining after velocity cleaning -- this is the final sample; (7) magnitude cut value; (8) radius limit for $\Qfit$/rms baseline; (9) inner radius for $\Qfit$/rms convergence test; (10) outer radius for $\Qfit$/rms convergence test; (11) $\Qfit$/rms percentile cut value.
\end{table*}

\subsection{Selection of bright stars}
\label{sect:select_bright}

As a result of two-body interactions, we expect that velocity dispersion will change with stellar mass, even if the cluster is not in full equipartition \citep[see e.g.][]{trenti2013}. Indeed, one of the huge advantages of the \textit{HST} PM catalogues that we have compiled is that we are able to get velocity measurements for stars on the main sequence; we are not limited to only the brightest (and most massive) stars, as is typically true from ground-based measurements. However, changes in dispersion with mass are beyond the scope of this paper. Here, we will restrict ourselves to studying the change of dispersion with spatial position in the cluster, so we need to trim our sample to include only stars inside a small range of mass. To do this, we select only the bright stars -- for this paper, we consider ``bright'' to be stars brighter than one magnitude below the main-sequence turnoff (MSTO).

To identify the MSTO, we bin all stars into magnitude bins of width 0.1 mag and then fit a Gaussian to the colour distribution in each bin to estimate the mean colour in the bin. Fitting a Gaussian is more robust in the presence of outliers than calculating a statistical mean. The bin with the bluest mean is considered to represent the turnoff\footnote{For some clusters, all parts of the colour-magnitude diagram (CMD) are well-populated and this simple algorithm can pick out the horizontal branch (HB) instead of following the trend of the main sequence (MS) or giant branch (GB) in a given magnitude bin. As the HB is bluewards of the MS and GB, this gives a false identification of the MSTO. To prevent such misidentification, we calculate the difference in mean colour between adjacent bins and check that the bins adjacent to the candidate MSTO have similar mean colours. If not, the bin with the next-bluest mean colour is considered. We find that this check is sufficient to ensure that we identify the correct MSTO.}.

\begin{figure}
    \centering
    \includegraphics[width=\linewidth]{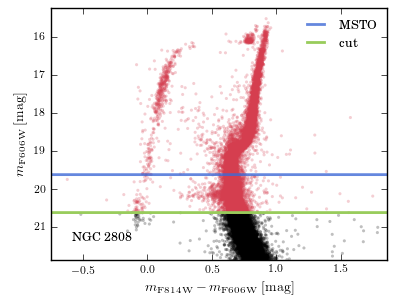}
    \caption{CMD illustrating bright star selection for NGC\,2808. The blue line marks the MSTO we identify using the algorithm described in the text; the green line represents the magnitude limit (1 mag below the MSTO) at which we cut to select only bright stars. Red points are stars that pass the cut, black points are stars that fail the cut.}
    \label{fig:select_bright}
\end{figure}

\autoref{fig:select_bright} shows a CMD illustrating the bright star selection for NGC\,2808. The blue line represents the identified MSTO magnitude and the green line marks the position of the bright-star selection cut at 1 mag below the MSTO. Red points show the stars we select for our bright sample and black points show the stars we discard. The cut magnitudes $M_\mathrm{cut}$ for each cluster are listed in \autoref{table:numbers}, along with the original number of stars $N_\mathrm{catalogue}$ in each catalogue and the number of stars $N_\mathrm{bright}$ that pass the magnitude selection.

\subsection{Data-quality selections}
\label{sect:quality_cuts}

Proper motions are estimated by determining the position of a star in a series of epochs and fitting a straight line to the positions as a function of time, thus the accuracy of the proper motions is dependent on the accuracy of the estimated positions. When a star suffers from blending -- that is, when the light from the star overlaps with a near neighbour -- those positions and, hence, the proper-motion estimate of the star may be subject to systematic errors.

To mitigate the effects of blending, we use several data-quality statistics reported in the cluster catalogues in order to select high-quality samples.  Crowded fields suffer more acutely from blending, so distant clusters and clusters with high central concentration will be the most affected. As such, there is no one-size-fits-all algorithm that can be applied. Some of the data-quality selections we describe below must be done on a cluster-by-cluster basis. We aim to be as consistent as possible with the treatment of each cluster, but in some cases, it is necessary to adapt our methods, as we will describe.

To begin, we consider the number of points used to determine the proper-motion fit. The cluster catalogues list both the total number of data points found for each star $\Nfound$ and the number of data points used for the final calculation of the proper motion $\Nused$. We assume that a star with $\Nused << \Nfound$ had many poor-quality measurements and that the remaining measurements may not be reliable. We cut all stars with $\Nuf < 0.8$.

The catalogues also provide $\chi^2_\mathrm{reduced}$ values separately for the $\mux$ and $\muy$ proper-motion fits. As these were straight-line fits, the number of degrees of freedom for each star is $D = \Nused - 2$. We calculate
\begin{equation}
    \alpha = F_\mathrm{D} \left( D \chi^2_\mathrm{reduced} \right)
\end{equation}
separately for both of the proper-motion directions, where $F_\mathrm{D}$ is the cumulative distribution function for a $\chi^2$ distribution with $D$ degrees of freedom. We cut all stars with $\alpha_\mathrm{x} > 0.99$ or $\alpha_\mathrm{y} > 0.99$. The number of stars $N_\mathrm{quality1}$ that pass the $\Nuf$ and $\chi^2$ selection process for each cluster are given in \autoref{table:numbers}.

Finally, we consider the quality of the point-spread function (PSF) fits to the stellar profiles. These are provided via quality-of-fit ($\Qfit$) values \citep[defined in][]{anderson2008} in all cluster catalogues except that for \wcen, which instead lists the root mean square (rms) of the magnitudes; both the $\Qfit$ and rms statistics behave similarly and may be treated in the same way.

We wish to excise all stars where the stellar profile has been poorly fit by the PSF model. In general, the quality of the PSF fit changes as a function of magnitude -- the stellar profiles of brighter stars are better approximated by a PSF than fainter stars -- so the cuts we make must also change with magnitude. As we have cut to select only bright stars, the changes with magnitude are small in most cases, but they must still be accounted for. PSF quality also changes with distance from the cluster centre as high crowding near the centre leads to poor PSF fits.

\begin{figure*}
    \centering
    \includegraphics[width=0.45\linewidth]{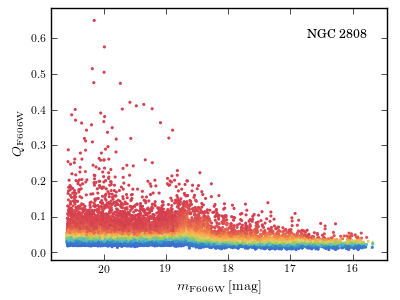}
    \includegraphics[width=0.45\linewidth]{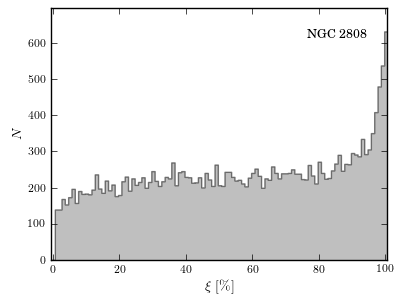}
    \includegraphics[width=0.45\linewidth]{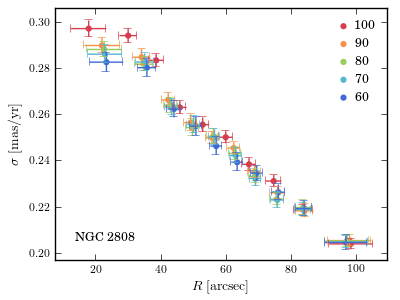}
    \includegraphics[width=0.45\linewidth]{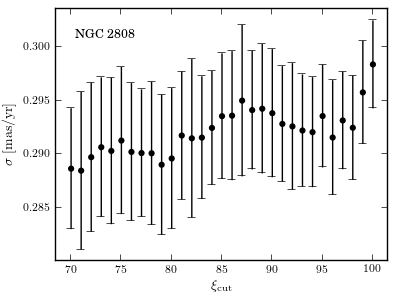}
    \caption{Illustration of of the $\Qfit$-cleaning process for NGC~2808. Top left: $\Qfit$ as a function of magnitude, here the stars are coloured according to their $\Qfit$ percentile from red (100$^\mathrm{th}$ percentile) to blue (0$^\mathrm{th}$ percentile). Top right: a histogram of $\Qfit$ percentiles. There is a sharp increase in the number of stars with percentiles of $\sim 85$ and higher. We use only stars outside of some limiting radius (40~arcsec, in this case) to calculate percentile values, so we attribute these high percentiles to poorly-fitted PSFs for stars near the centre. Bottom left: radial velocity-dispersion profiles for different $\Qfit$ percentile $\xi$ cuts. In the outer parts, the dispersion does not change for different percentile cuts, however, in the centre, the cuts do affect the profile. Bottom right: dispersion within the central 20~arcsec as a function of $\xi$ cut value. The dispersion decreases with decreasing $\xi$ cut until $\xi \sim 80$ and then levels off, so we adopt a cut of $\xi = 80$ for this cluster.}
    \label{fig:qfits}
\end{figure*}

For each star in the cluster, we identify the nearest 100 stars in magnitude that lie outside of some limiting distance $R_\mathrm{lim}$ from the centre and then calculate in which percentile $\xi$ the target star lies relative to its neighbours. We impose this radial distance limit in order to reduce the effects of crowding in the central regions; in this way we use predominantly isolated stars (with good PSF fits) as a baseline for comparison. For most clusters, we found that using an $R_\mathrm{lim}$ of 40~arcsec was sufficient, however, for the most crowded clusters in our sample (NGC 6388, NGC 6441, NGC 7078), we required an $R_\mathrm{lim}$ of 60~arcsec. For NGC 7099, which has the smallest field of view of all the clusters in our sample, we do not have enough stars to use any radial distance limit, so the $\Qfit$ percentiles were calculated using all stars. The limiting distances used for each cluster are listed in \autoref{table:numbers}.  We illustrate the results of this process for NGC\,2808 in \autoref{fig:qfits}. The top-left panel shows $\Qfit$ versus magnitude with the stars coloured according to their $\xi$ values, from the $\xi = 0$ in blue to $\xi = 100$ in red. The top-right panel shows a histogram of the $\xi$ values.

To determine at which percentile we should cut to ensure that we are left with only reliably-measured stars, we need to look at the dispersion profiles of the stars using different cuts (after appropriate velocity cleaning, which we discuss further below). The bottom-left panel of \autoref{fig:qfits} shows dispersion as a function of radius for different $\xi$ cuts. We consider that a cut is sufficient when further cuts make no change to the dispersion profile.

In general, we find that the profiles in the outer regions converge more quickly than those in the inner regions, so it is the inner regions we need to consider most carefully when deciding where to cut. As such, we select stars within a radius range ($R_\mathrm{in}$ to $R_\mathrm{out}$) at or near the centre, and calculate the dispersion of those stars for different $\xi$ cuts (again after appropriate velocity cleaning). For most clusters, we use $R_\mathrm{in} = 0$~arcsec and $R_\mathrm{out} = 20$~arcsec. Given the large distance (26.5~kpc) of NGC\,6715 and the high central concentration of NGC\,6624, we find better results using $R_\mathrm{out} = 10$~arcsec for these clusters; we also use an $R_\mathrm{out}$ of 10 arcsec for NGC\,7099 due to its small sample size. Further, there are a few clusters for which there are no faint stars at the very centre; for these clusters, we find that the PSF fits tend to be better for bright stars at the centre than for faint stars just outside of the centre. As such, for these clusters, we shift the radial range outwards: for NGC\,7078, we use $R_\mathrm{in} = 10$~arcsec and $R_\mathrm{out} = 30$~arcsec, and for NGC\,6388 and NGC\,6441, we use $R_\mathrm{in} = 20$~arcsec and $R_\mathrm{out} = 40$~arcsec.

The bottom-right panel of \autoref{fig:qfits} shows the change in dispersion as a function of $\xi$ cut value. For this example, the dispersion decreases from $\xi = 100$ to $\xi = 82$ and then levels off, so we adopt a cut value of $\xi = 82$ for this cluster. The radial range used for the convergence tests and the cut values adopted for each cluster are given in \autoref{table:numbers}, along with the number of stars $N_\mathrm{quality2}$ that pass the $\Qfit$/rms selection process.

\subsection{Velocity cleaning}
\label{sect:velocity_cleaning}

Outliers in a velocity distribution, largely due to contaminating populations, can bias dispersion estimates. While some modelling techniques can account for outliers directly \citep[e.g.][]{watkins2013}, here we will use simple models that require outliers to be excised. We need to take care with their removal, as cuts that are too harsh will remove cluster stars as well as outliers and this can also bias dispersion estimates.

Correct characterisation of velocity uncertainties is also important. Using a sample of ground-based proper motions for \wcen, \citet{vandeven2006} found that including stars with large uncertainties artificially increased their dispersion estimates and \citet{watkins2013} neatly showed how including these stars can return incorrect best-fit models. The problem with these stars was not so much that their uncertainties were large, as any reasonable dispersion estimator or modelling technique will take the size of uncertainties into account, but that the uncertainties had been underestimated, thus giving more weight to measurements than was warranted by their (poor) quality. That said, stars with uncertainties larger than approximately half of the cluster dispersion will contribute very little to our understanding of the cluster as the noise is larger than the signal. To ensure that we have a sample of stars with reliable uncertainties and reasonable signal-to-noise, we need to remove stars with large uncertainties.

In general, as we have already discussed, we expect velocity dispersion to change with stellar mass due to energy equipartition, but recall, we have already selected only bright stars in order to limit the mass range of our sample, so this effect may be ignored. However, we also expect that the velocity dispersion will be highest at the centre of a cluster and will decrease with radius, so if we make cuts that depend on the velocity dispersion, these cuts must be made as a function of radius.

To perform velocity cleaning, we bin stars in radius and then, in each bin, we 3-sigma clip to remove outliers and use the clipped sample to calculate a velocity dispersion. A third-order, flat-centre, monotonic, decreasing polynomial (see \autoref{sect:polyfit}) is then fitted to the binned dispersions and stars with errors larger than half of the fitted polynomial are removed. This method uses `bad' stars to estimate the dispersions, so we iterate three times to ensure the sample has been cleaned well.

\begin{figure}
    \centering
    \includegraphics[width=\linewidth]{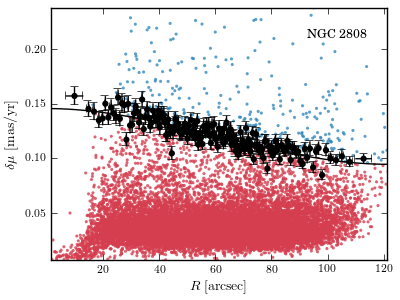}
    \caption{Illustration of the velocity error cuts for NGC\,2808. Black points show the 0.5 $\times$ the dispersion profile calculated from the data and the black line shows a third-order, flat-centre, monotonic, decreasing polynomial fit to the points. The red points are those stars below the line that pass the error cut and the blue points are the stars above the line that fail the error cut. We iterate this process three times where, in successive iterations, only stars passing the previous error cut are used to calculate the dispersion profile for determination of the new error cut. This graph shows the profile and cut status of the stars after the third iteration.}
    \label{fig:clean_velocity}
\end{figure}

\autoref{fig:clean_velocity} shows the results of the velocity cleaning at the end of the third and final iteration; the small points show the proper motion error $\delta \mu$ for each star as a function of radial position in the cluster. Red points are stars that survive the error cut, blue points are stars that removed by the error cut. The large black points show 0.5$\sigma$ for each of the radial bins and the black line shows the fitted polynomial used to make the cut. The number of stars $N_\mathrm{final}$ remaining in the final cleaned dataset for each cluster, after the error cuts and sigma clipping, is given in \autoref{table:numbers}. Note that we typically use only $\sim$15$\%$ of the catalogue stars for our present analysis. This is conservative, and does not mean that the remaining stars could not also be used to shed new light on cluster kinematics.


\section{Kinematics}
\label{sect:kinematics}

\subsection{Velocity dispersion estimates}
\label{sect:maxlh}

Throughout this paper, we assume that velocity distributions are Gaussian with a mean velocity $\overline{v}$ and a velocity dispersion $\sigma$.  We use a maximum-likelihood method to estimate the kinematic properties $\left( \overline{v}, \sigma \right)$ for a given set of stars.

Consider a sample of $N$ stars, where the $i$th star has an observed velocity $\vi$ with uncertainty $\dvi$. Then for some trial mean velocity $\vtrial$ and dispersion $\strial$, we ask what is the likelihood of observing the $i$th star:
\begin{equation}
    p \left( \vi \, \left| \vtrial, \strial, \dvi \right. \right) =
        \frac{1}{\sqrt{2 \pi \sigma'^2}}
        \exp{\left[ \frac{- \left( \vi - \vtrial \right)^2}
        {2 \sigma'^2} \right]}
\end{equation}
where
\begin{equation}
    \sigma' = \sqrt{\strial^2 + \dvi^2}
\end{equation}
Then the likelihood of observing the entire sample is the product of all likelihoods for the individual stars:
\begin{equation}
    \mathcal{L}_\mathrm{trial} = \prod_{i=1}^{N} p \left( \vi \, \left| \vtrial, \strial, \dvi \right. \right).
\end{equation}
The best fitting $\left( \overline{v}, \sigma \right)$ are those $\left( \vtrial, \strial \right)$ that maximise this likelihood.

\citet{vandeven2006} showed in their Appendix A that, for datasets with no measurement uncertainties, maximum-likelihood estimators are biased and underestimate the true dispersion of the velocity distribution by a factor
\begin{equation}
    b \left( N \right) = \sqrt{\frac{2}{N}}
    \frac{\Gamma \left( \frac{N}{2} \right)}
    {\Gamma \left( \frac{N-1}{2} \right)}
\end{equation}
where $\Gamma$ is the gamma function. When measurement uncertainties are present, there is no analytical correction, however they showed that an approximation to the correction can be written such that the corrected dispersion is
\begin{equation}
    \sigma_\mathrm{corr} \simeq \frac{1}{b \left( N \right)}
    \sqrt{ \sigma^2 + \left[ 1 - b^2 \left( N \right) \right]
    \overline{\delta v^2} }
\end{equation}
where $\sigma$ is the dispersion estimate obtained from the maximum-likelihood evaluation and
\begin{equation}
    \overline{\delta v^2} = \frac{1}{N} \sum_{i=1}^{N} \dvi^2
\end{equation}
is the mean of the squared measurement uncertainties. The difference between $\sigma_\mathrm{corr}$ and $\sigma$ is typically of order $0.1\%$ for the clusters in our sample.

Finally, we need to evaluate the uncertainties on our velocity and dispersion estimates. To do this we use a Monte-Carlo technique. Recall that for the purposes of this discussion, we are assuming a dataset of $N$ stars. We draw $M$ sets of $N$ samples from our best model $\left( \overline{v}, \sigma \right)$ and estimate the velocity mean and dispersion for each of the $M$ draws using the same maximum-likelihood method described above. The uncertainty on the mean is given by the dispersion of the Monte-Carlo means and the uncertainty on the dispersion is given by the dispersion of the Monte-Carlo dispersions. As the bias correction described above is only approximate, it is possible that the estimated dispersion and the uncertainty on the dispersion still have some bias; we estimate (and correct for) this remaining bias using the Monte-Carlo draws.

\subsection{Fitting a flat-centre, monotonic, decreasing polynomial}
\label{sect:polyfit}

For the velocity cleaning and for the analysis of our one-dimensional dispersion profiles, we wish to fit a third- or fourth-order polynomial to the binned cluster kinematics. We require that these polynomials: 1) be flat at the centre; 2) be monotonic; and 3) decrease with radius.

The first condition is easily satisfied by setting the coefficient of the $R^1$ term to zero so that the derivative of the fitted polynomial is zero at the centre.  The second condition is satisfied by taking the derivative of the polynomial and requiring that there be no roots of the derivative in the region of interest.  If the second condition holds true, then the third condition may be satisfied by insisting that the highest dispersion is found at the centre.


\section{Results}
\label{sect:results}

Now that we have datasets for each cluster that are of high photometric and astrometric quality and have laid out our kinematical analysis tools, we can study the kinematics of each cluster. We begin with a traditional one-dimensional analysis and investigate changes in velocity dispersion as a function of radius. Then we move on to a more advanced two-dimensional spatial analysis that highlights both the size and quality of these exquisite datasets.

\subsection{One-dimensional binning}
\label{sect:kin1d}

For one-dimensional spatial binning, we need to collapse two-dimensional spatial information into one dimension; that is, we go from x and y coordinates on the plane of the sky to radial distances from the cluster centre. Then we can use fixed-width bins, with variable area and population; fixed-area bins, with variable width and population; or bins of equal population, with variable width and area.

There are a number of competing effects we may wish to consider here. Most importantly, the spatial coverage of the original catalogues is inhomogenous; this is due to the observed fields available for our proper-motion analysis and is further compounded by the catalogue construction process and by the cleaning process we subsequently employ in \autoref{sect:catalogues}. We also note that the surface number density of stars in the cluster will decrease with radius. Overall, we find that we have fewer stars near the centre and towards the edges of the datasets than we do at intermediate radii. Furthermore, in order to get reliable estimates of velocity dispersion, we require a certain minimum number of stars per bin; we will address this point further below. Finally, we wish to have reasonable spatial resolution, particularly in the centre. With these considerations in mind, we adopt a combined approach: we require bins to be approximately equally populated (with some minimum required population $\Nstar$) to account for the inhomogeneities, however, we enforce different requirements over different radial ranges in order to maximise spatial resolution.

Our one-dimensional binning proceeds as follows:
\begin{enumerate}
    \item Make a central bin using the innermost $\Nstar$ stars.
    \item Calculate the number of unbinned stars with $R \le 5$~arcsec. If there are more than $\Nstar$ stars, make up to 3 equally populated bins containing at least $\Nstar$ stars.
    \item Calculate the number of unbinned stars with $R \le 10$~arcsec. If there are more than $\Nstar$ stars, make up to 3 equally populated bins containing at least $\Nstar$ stars.
    \item Calculate the number of unbinned stars with $R \le 20$~arcsec. If there are more than $\Nstar$ stars, make up to 3 equally populated bins containing at least $\Nstar$ stars.
    \item Calculate the number of unbinned stars remaining. If there are more than $\Nstar$ stars, make up to 20 equally populated bins containing at least $\Nstar$ stars.
\end{enumerate}

In each bin, we estimate the mean velocity $\vbar$ and velocity dispersion $\sigma$ of the combined radial and tangential proper-motion distributions, along with their uncertainties, using the maximum-likelihood technique described in \autoref{sect:maxlh}. We also calculate the radial ($\vr$ and $\sr$) and tangential ($\vt$ and $\st$) kinematics separately, and then use $\st / \sr$ as an indication of the anisotropy in the bin. We use the mean and standard deviation of the radial distances from the cluster centre to indicate the position of the bin and its uncertainty.

The number of stars we require per bin depends on the results we want to get out of the analysis and the particular science we want to do with those results. In this paper, our primary goal is simply to study the kinematical properties of the clusters; in-depth modelling of the underlying physics will be reserved for future papers. Further, we wish only to obtain first and second moments (mean and dispersion) of the velocity distribution for a set of stars, which requires fewer stars than would estimation of higher order moments. A fractional error on the dispersion of $\sim 10 \%$ will be sufficient for our purposes. Recall that fractional error $f$ on the dispersion is
\begin{equation}
    f = \frac{1}{\sqrt{2 N_\mathrm{v}}},
\end{equation}
where $N_\mathrm{v}$ is the number of velocity measurements used to calculate the dispersion. As we calculate dispersions using the combined radial and tangential proper motions, for a bin containing $\Nstar$ stars, $N_\mathrm{v} = 2 \Nstar$. Thus, if we require that $f = 0.1$, we find that $\Nstar = 25$. This will imply a slightly higher maximum fractional error on the anisotropy estimates of $f = 0.14$, which is still acceptable for our purposes.

\begin{figure}
    \centering
    \includegraphics[width=\linewidth]{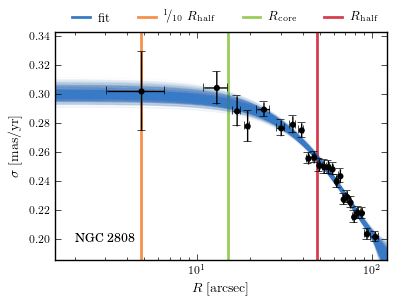}
    \includegraphics[width=\linewidth]{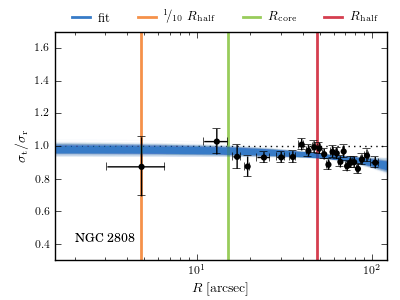}
    \caption{1D velocity profiles as a function of radius for NGC\,2808. The black points are the binned quantities that we estimate from the data. The green lines mark the core radius and the red and orange lines mark the half-light radius and 0.1 x the half-light radius respectively. Both the core radius and the half-light radius were taken from \citetalias{harris1996} and are given in \autoref{table:basics}. Top: Velocity dispersion profile. The blue lines are draws from an MCMC fit of a fourth-order, monotonic, decreasing polynomial that we force to be flat at the centre. Bottom: Velocity anisotropy profile. The blue lines are draws from an MCMC straight-line fit; the fits are made to be linear with radius and, hence, appear curved in the figure. The dotted line indicates isotropy at $\st / \sr = 1$.}
    \label{fig:kin1d}
\end{figure}

The upper panel of \autoref{fig:kin1d} shows the combined radial and tangential dispersion profile as function of radius for NGC\,2808. The black points are the binned dispersions that we estimate from the data. We see that the dispersion is highest at the centre and falls off with radius, as expected. We show the dispersion profiles for all other clusters in the left columns of Figures~\ref{fig:kin1}--\ref{fig:kin6}.

In general, scatter in the dispersion profiles increases towards the centre of the clusters due to the small number of stars per bin near the centres, as  reflected by the increasing error bars. However, for NGC\,7078 (M\,15), we also see a clear dip in the dispersion profile at the centre of the cluster. We believe that this is not intrinsic to the cluster, but is merely an artefact due to magnitude-dependent incompleteness (in the most crowded central regions only the brightest stars survive our quality cuts, so that the dispersion measurement pertains to a brighter magnitude range than at larger radii). We note that the dispersion profile for this cluster using RGB and SGB stars presented in \citetalias{bellini2014} did not see the same central dip. In that case, no $\Qfit$ cuts were made before the calculation of that profile, which may explain the difference. The results for NGC 7078 highlight that, for any detailed dynamical modelling of these datasets, it is important that both incompleteness and the magnitude dependence of the kinematics be taken into account. NGC 7078 is probably the cluster in our sample that is most affected by these issues, due to both its high central density, and its large distance. We do not generally see statistically significant dips in the central dispersion profiles for other clusters, so there is no reason to be suspicious of the reported dispersion profiles in general.

Another peculiar feature seen in the dispersion profiles of a few clusters is an upturn in the outer regions. In NGC\,288 and NGC\,362, we attribute this to edge effects, and, in NGC\,6535, to the small number of stars in the dataset. We must also bear in mind that some clusters are rather inhomogeneous in their spatial coverage, especially in the outer parts. This means that different datasets are used to derive proper motions at different radii. This has the potential to introduce small systematics, which may also explain other, more subtle bumps and wiggles observed in some of the profiles.

Near to the cluster tidal radii, we would also expect interesting tidal effects to manifest in the dispersion profiles. Tidal radii are notoriously difficult to estimate. \citetalias{harris1996} provides core radii and King concentration parameters from which tidal radii can, in principle, be estimated, but cautions that tidal radii calculated in this way are not reliable. Nevertheless, we can use the King profile estimates as approximate lower limits on the tidal radii as we know that King profiles will tend to underestimate tidal radii \citepalias[e.g.][]{mclaughlin2005}.\footnote{Even this is not entirely correct, as we know that the tidal radius of a cluster depends on the orbit of the cluster through the tidal field of the Milky Way and that the tidal field can affect clusters well within the tidal radius \citep[e.g.][]{webb2013, moreno2014, kennedy2014}. Nevertheless, we do believe that our data is sufficiently far in to be unaffected by the tidal field.} We find that these (lower bounds on the) tidal radii exceed the limits of our data for all clusters. For NGC\,288 and NGC\,362, the tidal radii are approximately a factor of 8 and a factor of 6 larger than the radii of the outermost bins respectively, so we do not believe that the upturns in the outer regions of the dispersion profiles for these clusters are attributable to tidal effects.

\begin{table*}
    \centering
    \caption{One-dimensional kinematic data results.}
    \label{table:kin1d}
    \begin{tabular}{ccccccccc}
        \hline
        \hline
        & \multicolumn{2}{c}{at $R=0$} & \multicolumn{2}{c}{at $\Rcore$} & \multicolumn{2}{c}{at $\Rhalf$} & \multicolumn{2}{c}{at $0.1 \Rhalf$} \\
        Cluster & $\sigma$ & $\st/\sr$ & $\sigma$ & $\st/\sr$ & $\sigma$ & $\st/\sr$ & $\sigma$ & $\st/\sr$ \\
        & (\masyr) & & (\masyr) & & (\masyr) & & (\masyr) & \\
        (1) & (2) & (3) & (4) & (5) & (6) & (7) & (8) & (9) \\
        \hline
        NGC\,104  & 0.573 $\pm$ 0.005 & 1.01 $\pm$ 0.01 & 0.570 $\pm$ 0.004 & 1.00 $\pm$ 0.01 & \dots             & \dots           & 0.571 $\pm$ 0.004 & 1.00 $\pm$ 0.01 \\
        NGC\,288  & 0.076 $\pm$ 0.003 & 1.00 $\pm$ 0.06 & 0.063 $\pm$ 0.001 & 0.95 $\pm$ 0.03 & \dots             & \dots           & 0.074 $\pm$ 0.002 & 0.99 $\pm$ 0.05 \\
        NGC\,362  & 0.198 $\pm$ 0.002 & 1.03 $\pm$ 0.02 & 0.195 $\pm$ 0.001 & 1.02 $\pm$ 0.01 & 0.169 $\pm$ 0.001 & 0.99 $\pm$ 0.01 & 0.197 $\pm$ 0.002 & 1.03 $\pm$ 0.01 \\
        NGC\,1851 & 0.183 $\pm$ 0.002 & 1.02 $\pm$ 0.02 & 0.182 $\pm$ 0.002 & 1.02 $\pm$ 0.02 & 0.158 $\pm$ 0.001 & 0.99 $\pm$ 0.01 & \dots             & \dots           \\
        NGC\,2808 & 0.301 $\pm$ 0.004 & 0.99 $\pm$ 0.02 & 0.295 $\pm$ 0.003 & 0.97 $\pm$ 0.01 & 0.257 $\pm$ 0.001 & 0.95 $\pm$ 0.01 & 0.300 $\pm$ 0.003 & 0.98 $\pm$ 0.01 \\
        NGC\,5139 & 0.767 $\pm$ 0.004 & 1.01 $\pm$ 0.01 & 0.640 $\pm$ 0.002 & 0.93 $\pm$ 0.00 & 0.557 $\pm$ 0.031 & 0.83 $\pm$ 0.01 & 0.754 $\pm$ 0.003 & 0.99 $\pm$ 0.01 \\
        NGC\,5904 & 0.224 $\pm$ 0.002 & 1.01 $\pm$ 0.02 & 0.217 $\pm$ 0.001 & 0.99 $\pm$ 0.01 & 0.187 $\pm$ 0.003 & 0.94 $\pm$ 0.02 & 0.223 $\pm$ 0.002 & 1.00 $\pm$ 0.02 \\
        NGC\,5927 & 0.168 $\pm$ 0.001 & 0.99 $\pm$ 0.01 & 0.163 $\pm$ 0.001 & 0.98 $\pm$ 0.01 & 0.150 $\pm$ 0.001 & 0.98 $\pm$ 0.01 & 0.168 $\pm$ 0.001 & 0.99 $\pm$ 0.01 \\
        NGC\,6266 & 0.502 $\pm$ 0.004 & 0.99 $\pm$ 0.01 & 0.493 $\pm$ 0.003 & 0.99 $\pm$ 0.01 & 0.419 $\pm$ 0.002 & 0.98 $\pm$ 0.01 & 0.500 $\pm$ 0.004 & 0.99 $\pm$ 0.01 \\
        NGC\,6341 & 0.211 $\pm$ 0.003 & 1.01 $\pm$ 0.02 & 0.206 $\pm$ 0.002 & 1.00 $\pm$ 0.02 & 0.174 $\pm$ 0.001 & 0.98 $\pm$ 0.01 & 0.210 $\pm$ 0.002 & 1.00 $\pm$ 0.02 \\
        NGC\,6362 & 0.108 $\pm$ 0.003 & 1.05 $\pm$ 0.06 & 0.100 $\pm$ 0.001 & 0.95 $\pm$ 0.03 & \dots             & \dots           & 0.107 $\pm$ 0.003 & 1.03 $\pm$ 0.05 \\
        NGC\,6388 & 0.307 $\pm$ 0.005 & 0.97 $\pm$ 0.02 & 0.305 $\pm$ 0.005 & 0.97 $\pm$ 0.02 & 0.282 $\pm$ 0.002 & 0.97 $\pm$ 0.01 & \dots             & \dots           \\
        NGC\,6397 & 0.449 $\pm$ 0.009 & 0.98 $\pm$ 0.03 & 0.448 $\pm$ 0.009 & 0.98 $\pm$ 0.03 & \dots             & \dots           & 0.443 $\pm$ 0.007 & 0.97 $\pm$ 0.03 \\
        NGC\,6441 & 0.277 $\pm$ 0.004 & 0.95 $\pm$ 0.02 & 0.276 $\pm$ 0.004 & 0.95 $\pm$ 0.02 & 0.260 $\pm$ 0.002 & 0.94 $\pm$ 0.01 & \dots             & \dots           \\
        NGC\,6535 & 0.100 $\pm$ 0.007 & 0.79 $\pm$ 0.12 & 0.095 $\pm$ 0.005 & 0.79 $\pm$ 0.08 & 0.086 $\pm$ 0.004 & 0.79 $\pm$ 0.06 & 0.099 $\pm$ 0.007 & 0.79 $\pm$ 0.11 \\
        NGC\,6624 & 0.192 $\pm$ 0.004 & 0.97 $\pm$ 0.04 & 0.191 $\pm$ 0.004 & 0.97 $\pm$ 0.04 & 0.140 $\pm$ 0.002 & 1.01 $\pm$ 0.02 & 0.190 $\pm$ 0.004 & 0.97 $\pm$ 0.04 \\
        NGC\,6656 & 0.596 $\pm$ 0.008 & 1.00 $\pm$ 0.02 & 0.542 $\pm$ 0.004 & 1.00 $\pm$ 0.01 & \dots             & \dots           & 0.586 $\pm$ 0.005 & 1.00 $\pm$ 0.01 \\
        NGC\,6681 & 0.153 $\pm$ 0.002 & 1.03 $\pm$ 0.02 & 0.152 $\pm$ 0.002 & 1.03 $\pm$ 0.02 & 0.117 $\pm$ 0.001 & 0.99 $\pm$ 0.01 & 0.152 $\pm$ 0.002 & 1.03 $\pm$ 0.02 \\
        NGC\,6715 & 0.152 $\pm$ 0.003 & 1.00 $\pm$ 0.03 & 0.150 $\pm$ 0.003 & 1.00 $\pm$ 0.03 & 0.100 $\pm$ 0.001 & 0.99 $\pm$ 0.02 & 0.151 $\pm$ 0.003 & 1.00 $\pm$ 0.03 \\
        NGC\,6752 & 0.413 $\pm$ 0.005 & 0.97 $\pm$ 0.02 & 0.411 $\pm$ 0.005 & 0.97 $\pm$ 0.02 & 0.310 $\pm$ 0.004 & 0.97 $\pm$ 0.02 & 0.410 $\pm$ 0.005 & 0.97 $\pm$ 0.02 \\
        NGC\,7078 & 0.234 $\pm$ 0.003 & 1.04 $\pm$ 0.02 & 0.232 $\pm$ 0.003 & 1.02 $\pm$ 0.02 & 0.181 $\pm$ 0.001 & 0.90 $\pm$ 0.01 & 0.233 $\pm$ 0.003 & 1.03 $\pm$ 0.02 \\
        NGC\,7099 & 0.304 $\pm$ 0.041 & 0.86 $\pm$ 0.22 & 0.278 $\pm$ 0.028 & 0.92 $\pm$ 0.16 & \dots             & \dots           & 0.246 $\pm$ 0.015 & 0.96 $\pm$ 0.12 \\
        \hline
    \end{tabular}
    
    \qquad 
    
    \textbf{Notes.} Columns: (1) cluster identification in the NGC catalogue; (2) central dispersion; (3) central anisotropy; (4) dispersion at core radius; (5) anisotropy at core radius; (6) dispersion at half-light radius; (7) anisotropy at half-light radius; (8) dispersion at 0.1x half-light radius; (9) anisotropy at 0.1x half-light radius. The core radii and half-light radii used are shown in \autoref{table:basics}.
\end{table*}

To each of the binned profiles, we fit a fourth-order, monotonic, decreasing polynomial, which we force to be flat at the centre\footnote{Globular-cluster dispersion profiles may not be flat near the centre (e.g. if there is an intermediate-mass black hole); we make this assumption here to conveniently describe the data over the radial range where it is available. The profiles should be extrapolated only with care.} (see \autoref{sect:polyfit}). To do this, we use the Markov Chain Monte Carlo (MCMC) package \textsc{emcee} developed by \citet{foremanmackey2013}, which is an implementation of the affine-invariant ensemble sampler by \citet{goodman2010}; this approach uses multiple trial points (walkers) at each step to efficiently explore the parameters space. We run our MCMC chain with 100 walkers for 1000 steps, and use the last 20 steps as the final (``post-burn'') sample. The polynomial fits from every second step of the final sample are shown as blue lines.

Finally, we use the core $\Rcore$ and half-light $\Rhalf$ radii given in \autoref{table:basics} and use the MCMC post-burn sample to estimate the dispersions at $\Rcore$, $\Rhalf$ and $0.1 \Rhalf$ (when each radius lies within the range covered by our datasets). These radii are marked as green ($\Rcore$), red ($\Rhalf$) and orange ($0.1 \Rhalf$) lines. The dispersion estimates at each radius, along with the central dispersion, are given in \autoref{table:kin1d}.

The lower panel of \autoref{fig:kin1d} shows the tangential-over-radial anisotropy profile as function of radius for NGC\,2808. The black points are the binned anisotropies we estimate from the data. The black dotted lines indicate isotropy at $\st / \sr = 1$. We see that the cluster velocities are isotropic ($\sr \sim \st$) at the centre and become mildly radially anisotropic ($\sr > \st$) with increasing radius. We observe this behaviour for most of the clusters in our sample; their anisotropy profiles are shown in the middle columns of Figures~\ref{fig:kin1}-\ref{fig:kin6}.

For the anisotropies, we fit each profile with a straight line using MCMC; we run the chain with 100 walkers for 500 steps, and again use the last 20 steps as the post-burn sample. The straight-line fits from every second step of the post-burn sample are shown as blue lines in the figures; note that the fits are linear with radius and so appear curved in the figure. As for the dispersions, we estimate the anisotropies at $\Rcore$, $\Rhalf$ and $0.1 \Rhalf$ and mark these radii in each figure as green, red and orange lines respectively. These anisotropy estimates and the central anisotropy are given in \autoref{table:kin1d}.

The full binned dispersion and anisotropy profiles for all clusters are provided in \autoref{sect:kin1d_profiles}.

\subsection{Two-dimensional binning in spatial coordinates}
\label{sect:kin2d}

In most cases, our cleaned datasets are large enough that we are not limited to binning only in one dimension; we have sufficient numbers of stars to bin in two dimensions while retaining reasonable resolution. We begin with two-dimensional spatial binning, that is, binning in x' and y' coordinates on the plane of the sky\footnote{Here the primes on the coordinates denote alignment with the cluster major and minor axes, with the exception of NGC\,288 for which no position angle was available in \citet{white1987} so the major axis could not be identified.}.

To bin in two dimensions, we could proceed by using bins of fixed size -- say in radius and azimuth -- or we could bin the stars so that the bins are approximately equally populated. Due to heterogenous spatial coverage of the individual datasets, as we discussed in \autoref{sect:kin1d}, we opt for the latter binning scheme. To do this, we use the Voronoi-binning algorithm described in \citep{cappellari2003}.\footnote{\citep{cappellari2003} kindly provided \textsc{IDL} code of their Voronoi-binning algorithm. We have converted this into \textsc{Python}; it is available at \url{http://github.com/lauralwatkins/voronoi}. We note that there is now a Python version available from Michele Cappellari also, however it was not available at the time our code was written.} We first bin our stars into a $30 \times 30$ grid of equally-spaced pixels in the x'--y' plane. The Voronoi algorithm then bins these pixels into irregular bins such that they contain approximately equal numbers of stars. (In fact, the algorithm seeks to equalise the signal-to-noise ratio in each bin; as we have discrete datasets, our `signal' is the number of stars in the bin $\Nstar$ and our `noise' is $\sqrt{\Nstar}$.) Due to the nature of the binning algorithm, there is moderate scatter on the population of each bin; as such, we use a higher target population of $\Nstar = 50$ for this analysis.

Due to the very small sample sizes of the NGC\,6535 and NGC\,7099 datasets, we are forced to make some adjustments to this procedure for these clusters: we use a much coarser $10 \times 10$ grid and use a target population of $\Nstar = 10$. For completeness, we include their results, but advise that they be considered with caution.

\begin{figure*}
    \centering
    \includegraphics[width=0.33\linewidth]{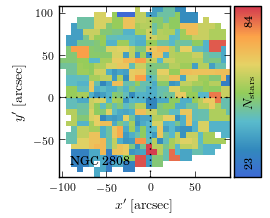}
    \includegraphics[width=0.33\linewidth]{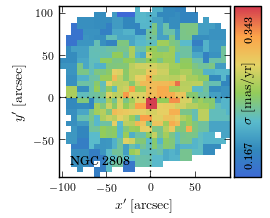}
    \includegraphics[width=0.33\linewidth]{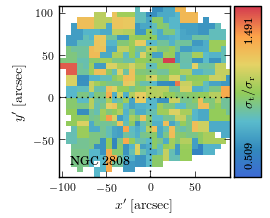}
    \caption{2D maps for NGC~2808 as a function of spatial coordinates. Stars were binned into equally-spaced pixels and the pixels were binned using a Voronoi algorithm \citep{cappellari2003} in order to achieve approximately equally populated bins. Each pixel is coloured according a particular quantity calculated using all stars in the bin to which the pixel belongs from red (high) to blue (low), as indicated by the colour bars. The dotted lines mark $x'=0$ and $y'=0$; their intersection marks the adopted cluster centre for this study. Left: Number of stars per bin. The spatial coverage of the sample is fairly homogeneous, with a slight underdensity at the centre. Middle: Velocity dispersion. Note that the dispersion is highest at the centre and falls off with radius. We do not see evidence of significant flattening of the velocity distribution. Right: Velocity anisotropy. Here the colour scale is symmetric such that $\st/\sr = 1$ (isotropy) is shown in green. We observed a change in anisotropy with radius in the 1D profile for this cluster but we do not see it here; we believe that the effect is too small to be seen against pixel-to-pixel variations.}
    \label{fig:kin2d_xy}
\end{figure*}

To illustrate the nature of the Voronoi binning, the left panel of \autoref{fig:kin2d_xy} shows a pixel map for NGC\,2808; each pixel is coloured according to the number of stars in the bin to which that pixel belongs, from red (high) to blue (low). In this case, the most populous bin contains 84 stars and the least populous bin contains 23 stars. The dotted lines mark $x'=0$ and $y'=0$ and their intersection marks the adopted centre of the cluster \citepalias[see Table 1 of][]{bellini2014}. Overall, the spatial coverage of the stars is fairly homogeneous, although we do note that there is a slight underdensity of stars at the centre due to selection effects.

Now we proceed as we did for the one-dimensional analysis. In each bin, we estimate the mean velocity $\vbar$ and velocity dispersion $\sigma$ of the combined radial and tangential proper-motion distributions, using the maximum-likelihood technique described in \autoref{sect:maxlh}. We also calculate the radial ($\vr$ and $\sr$) and tangential ($\vt$ and $\st$) kinematics separately, and then use $\st / \sr$ as an indication of the anisotropy in the bin.

\begin{figure*}
    \centering
    
    \includegraphics[width=0.33\linewidth]{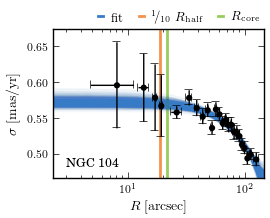}
    \includegraphics[width=0.33\linewidth]{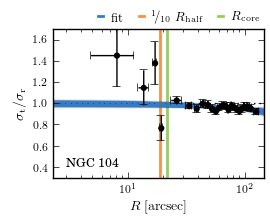}
    \includegraphics[width=0.33\linewidth]{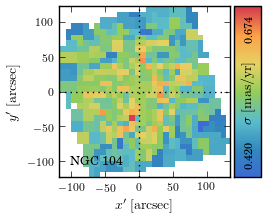}
    
    \includegraphics[width=0.33\linewidth]{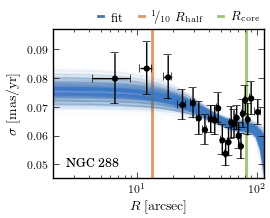}
    \includegraphics[width=0.33\linewidth]{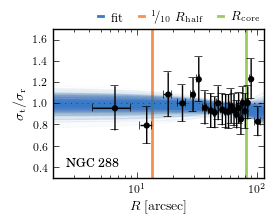}
    \includegraphics[width=0.33\linewidth]{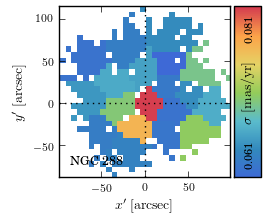}
    
    \includegraphics[width=0.33\linewidth]{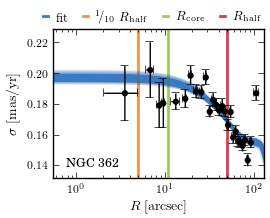}
    \includegraphics[width=0.33\linewidth]{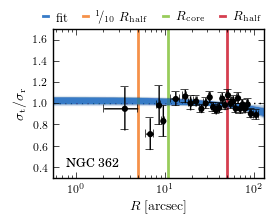}
    \includegraphics[width=0.33\linewidth]{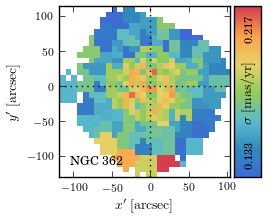}
    
    \includegraphics[width=0.33\linewidth]{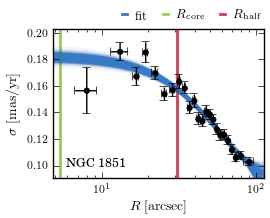}
    \includegraphics[width=0.33\linewidth]{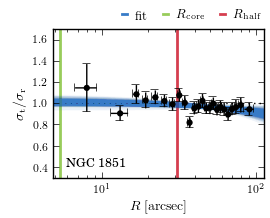}
    \includegraphics[width=0.33\linewidth]{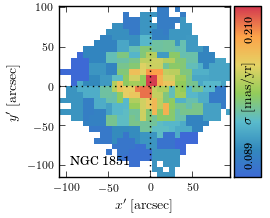}
    
    \caption{Kinematic results for NGC\,104, NGC\,288, NGC\,362 and NGC\,1851. Left: 1D velocity dispersion profiles, similar to the top panel of \autoref{fig:kin1d}. Middle: 1D velocity anisotropy profiles, similar to the bottom panel of \autoref{fig:kin1d}. Right: 2D velocity dispersion maps, similar to the middle panel of \autoref{fig:kin2d_xy}.}
    \label{fig:kin1}
\end{figure*}

\begin{figure*}
    \centering
    
    \includegraphics[width=0.33\linewidth]{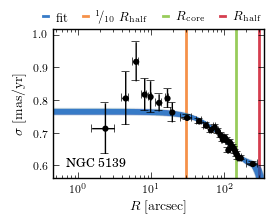}
    \includegraphics[width=0.33\linewidth]{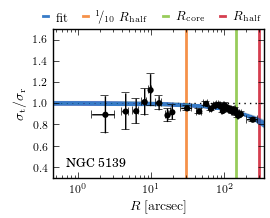}
    \includegraphics[width=0.33\linewidth]{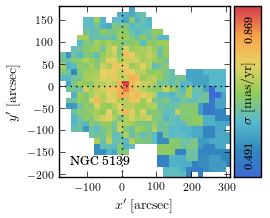}
    
    \includegraphics[width=0.33\linewidth]{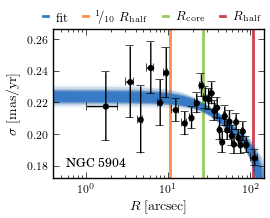}
    \includegraphics[width=0.33\linewidth]{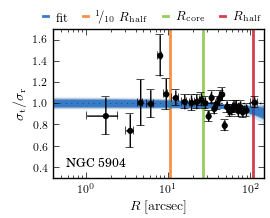}
    \includegraphics[width=0.33\linewidth]{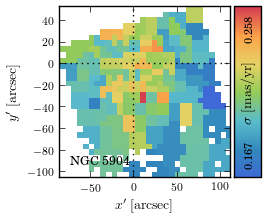}
    
    \includegraphics[width=0.33\linewidth]{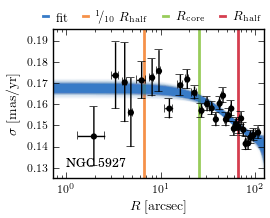}
    \includegraphics[width=0.33\linewidth]{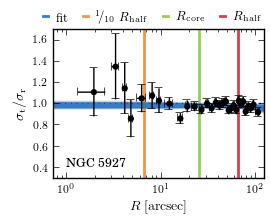}
    \includegraphics[width=0.33\linewidth]{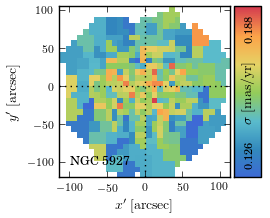}
    
    \includegraphics[width=0.33\linewidth]{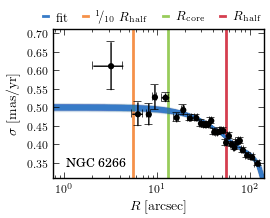}
    \includegraphics[width=0.33\linewidth]{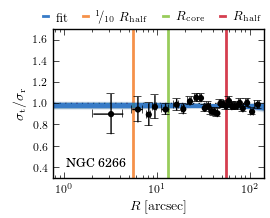}
    \includegraphics[width=0.33\linewidth]{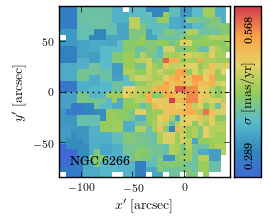}
    
    \caption{Similar to \autoref{fig:kin1} for NGC\,5139, NGC\,5904, NGC\,5927 and NGC\,6266.}
    \label{fig:kin2}
\end{figure*}

\begin{figure*}
    \centering
    
    \includegraphics[width=0.33\linewidth]{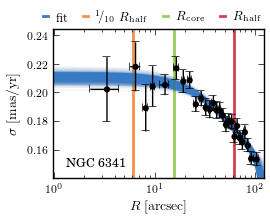}
    \includegraphics[width=0.33\linewidth]{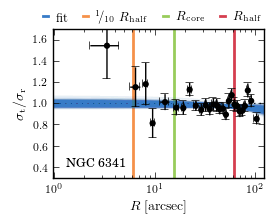}
    \includegraphics[width=0.33\linewidth]{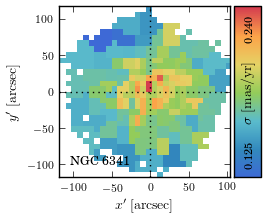}
    
    \includegraphics[width=0.33\linewidth]{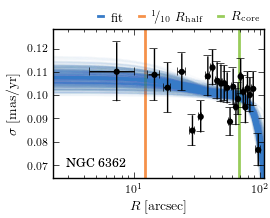}
    \includegraphics[width=0.33\linewidth]{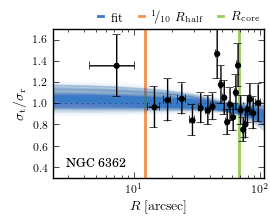}
    \includegraphics[width=0.33\linewidth]{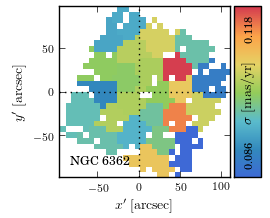}
    
    \includegraphics[width=0.33\linewidth]{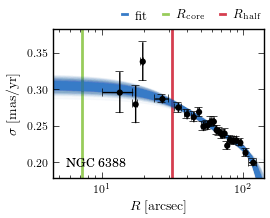}
    \includegraphics[width=0.33\linewidth]{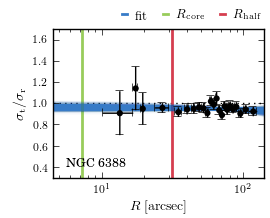}
    \includegraphics[width=0.33\linewidth]{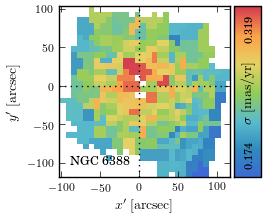}
    
    \includegraphics[width=0.33\linewidth]{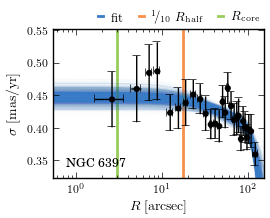}
    \includegraphics[width=0.33\linewidth]{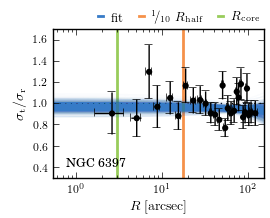}
    \includegraphics[width=0.33\linewidth]{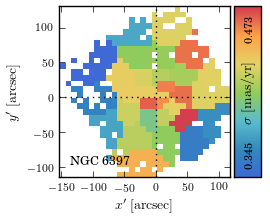}
    
    \caption{Similar to \autoref{fig:kin1} for NGC\,6341, NGC\,6362, NGC\,6388 and NGC\,6397.}
    \label{fig:kin3}
\end{figure*}

\begin{figure*}
    \centering
    
    \includegraphics[width=0.33\linewidth]{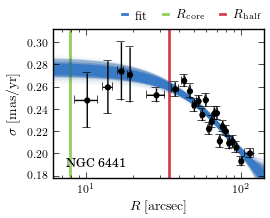}
    \includegraphics[width=0.33\linewidth]{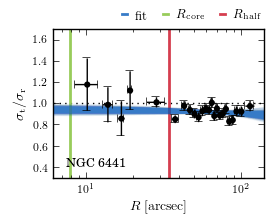}
    \includegraphics[width=0.33\linewidth]{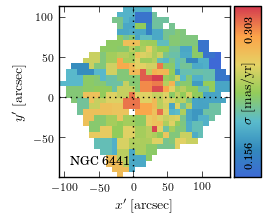}
    
    \includegraphics[width=0.33\linewidth]{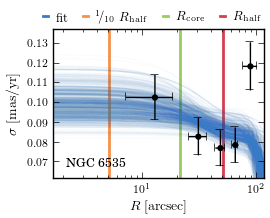}
    \includegraphics[width=0.33\linewidth]{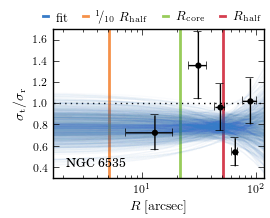}
    \includegraphics[width=0.33\linewidth]{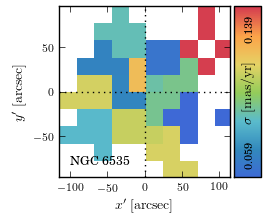}
    
    \includegraphics[width=0.33\linewidth]{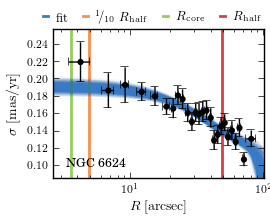}
    \includegraphics[width=0.33\linewidth]{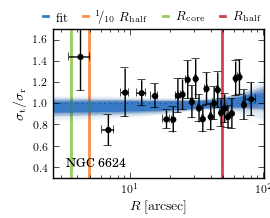}
    \includegraphics[width=0.33\linewidth]{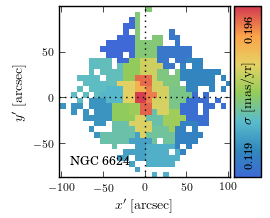}
    
    \includegraphics[width=0.33\linewidth]{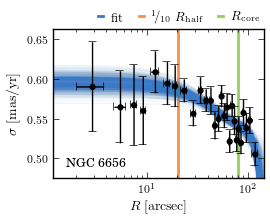}
    \includegraphics[width=0.33\linewidth]{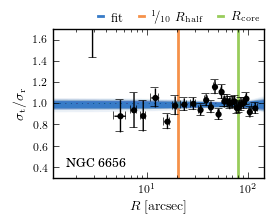}
    \includegraphics[width=0.33\linewidth]{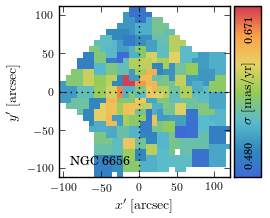}
    
    \caption{Similar to \autoref{fig:kin1} for NGC\,6441, NGC\,6535, NGC\,6624 and NGC\,6656.}
    \label{fig:kin4}
\end{figure*}

\begin{figure*}
    \centering
    
    \includegraphics[width=0.33\linewidth]{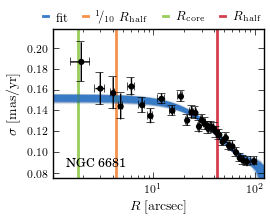}
    \includegraphics[width=0.33\linewidth]{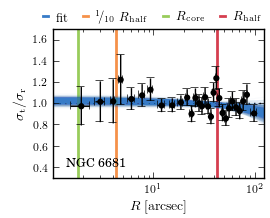}
    \includegraphics[width=0.33\linewidth]{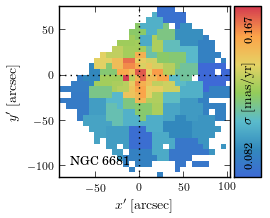}
    
    \includegraphics[width=0.33\linewidth]{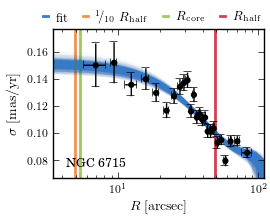}
    \includegraphics[width=0.33\linewidth]{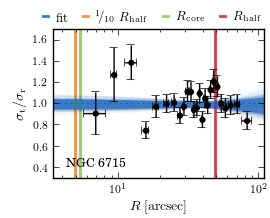}
    \includegraphics[width=0.33\linewidth]{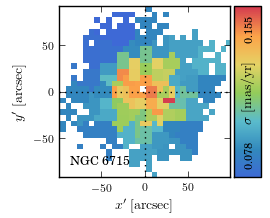}
    
    \includegraphics[width=0.33\linewidth]{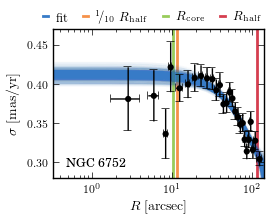}
    \includegraphics[width=0.33\linewidth]{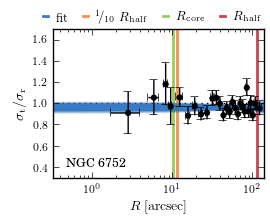}
    \includegraphics[width=0.33\linewidth]{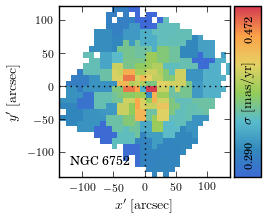}
    
    \includegraphics[width=0.33\linewidth]{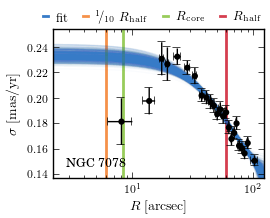}
    \includegraphics[width=0.33\linewidth]{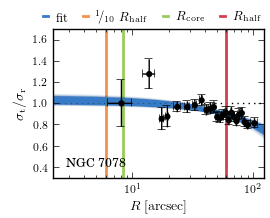}
    \includegraphics[width=0.33\linewidth]{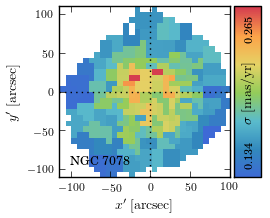}
    
    \caption{Similar to \autoref{fig:kin1} for NGC\,6681, NGC\,6715, NGC\,6752 and NGC\,7078.}
    \label{fig:kin5}
\end{figure*}

\begin{figure*}
    \centering
    
    \includegraphics[width=0.33\linewidth]{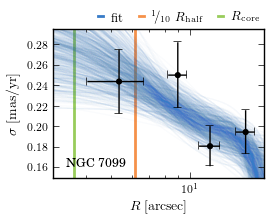}
    \includegraphics[width=0.33\linewidth]{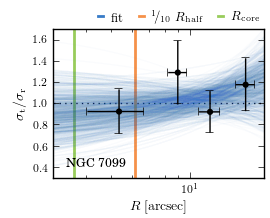}
    \includegraphics[width=0.33\linewidth]{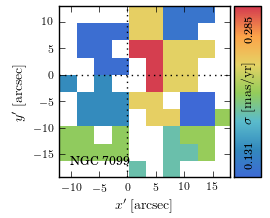}
    
    \caption{Similar to \autoref{fig:kin1} for NGC\,7099.}
    \label{fig:kin6}
\end{figure*}

The middle panel of \autoref{fig:kin2d_xy} shows the resulting velocity dispersion map for NGC\,2808. Each pixel is coloured according to the dispersion of the bin to which the pixel belongs from red (high) to blue (low). We can clearly see that the dispersion is highest at the centre of the cluster and falls off with radius, as expected. We also note that the dispersion profile is round, that is, it does not appear to fall off faster in any particular direction, so we see no evidence of significant flattening in this cluster. We show the 2D dispersion maps for the rest of our cluster sample in the right columns of Figures~\ref{fig:kin1}--\ref{fig:kin6}.

The right panel of \autoref{fig:kin2d_xy} shows the velocity anisotropy map for NGC\,2808. Each pixel is coloured according to the $\st/\sr$ anisotropy in the bin to which the pixel belongs from red (high) to blue (low). Here the colour scale is symmetric so that $\st/\sr = 1$ (isotropy) is shown in green. We are unable to discern any spatial patterns here. Recall that in the 1D anisotropy profiles, we saw that NGC\,2808 is nearly isotropic at the centre and becomes mildly radially anisotropic towards the outer parts of the cluster; however here, it seems that trend is washed out in the pixel-to-pixel noise. This is true for all clusters in our sample, so we do not include further 2D anisotropy profiles as we do not believe that any meaningful information can be taken from them.

\subsection{Two-dimensional binning in radius and magnitude}
\label{sect:kin2d_rmag}

\begin{figure*}
    \centering
    \includegraphics[width=0.33\linewidth]{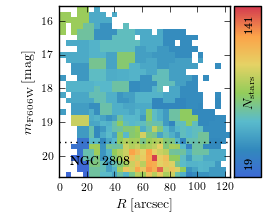}
    \includegraphics[width=0.33\linewidth]{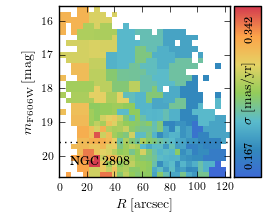}
    \includegraphics[width=0.33\linewidth]{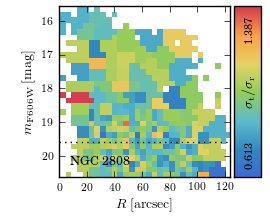}
    \caption{2D maps for NGC~2808 as a function of radius and magnitude. Stars were binned into equally-spaced pixels and the pixels were binned using a Voronoi algorithm \citep{cappellari2003} in order to achieve approximately equally populated bins. Each pixel is coloured according a particular quantity calculated using all stars in the bin to which the pixel belongs from red (high) to blue (low), as indicated by the colour bars. The dotted lines mark the position of the identified MSTO. Left: Number of stars per bin. Note that the inhomogeneous sample in the radius-magnitude plane, with very stars in the central 10~arcsec and an overabundance of faint stars at intermediate radii. Middle: Velocity dispersion. Dispersion clearly decreases with radius, however, also note that dispersion does decrease with increasing brightness (mass), albeit to a lesser extent, highlighting the importance of fully accounting for both stellar mass and position in the cluster in dynamical modelling studies. Right: Velocity anisotropy. Here the colour scale is symmetric such that $\st/\sr = 1$ (isotropy) is shown in green. We observed a change in anisotropy with radius in the 1D profile for NGC\,2808 but we do not see it here; we believe that the effect is too small to be seen against pixel-to-pixel variations.}
    \label{fig:kin2d_rmag}
\end{figure*}

In \autoref{sect:select_bright}, we imposed a brightness limit on our catalogues in order to minimise the mass range of the included stars. We did this because we expect that velocity dispersions will change with both stellar mass and the position of a star in the cluster, and we wish to address only the latter in this paper. In order to test the efficacy of the magnitude cut, we repeat our 2D analysis for NGC\,2808, but now we bin in radius and magnitude, which we use as a proxy for mass. As before, we split the radius-magnitude plane into a grid of $30 \times 30$ pixels and use the Voronoi algorithm with $\Nstar = 50$ to bin the pixels, then we calculate the kinematics for each bin. We show the results in \autoref{fig:kin2d_rmag}.

In the left panel, the pixels are coloured according to the number of stars in the bin to which the pixel belongs from red (high) to blue (low). The distribution of stars in the radius-magnitude plane is highly inhomogeneous. We see that the majority of our stars are (comparatively) faint -- a natural consequence of the luminosity function -- and are found at intermediate radii. Conversely, we have very few faint stars within the central 10~arcsec, although we note that there are few bright stars in that region also. The reason for this paucity of stars is two-fold: first, the number of stars near the centre will be small as the area is small (despite the increase in number density we expect near the centre); and second, crowding will be most problematic near the centre so this region will have been heavily impacted by our photometric and kinematic quality cuts. As crowding is a more serious issue for faint stars than for bright stars, this second consideration also explains why we see fewer faint stars than bright stars near the centre.

In the middle panel of \autoref{fig:kin2d_rmag}, the pixels are coloured according to the velocity dispersion of the stars in the bin to which the pixel belongs from red (high) to blue (low). We see here that, for all magnitudes, the velocity dispersions are highest near the centre and fall off with radius, just as we observed in the 1D dispersion profile. We also note that there is a small change in dispersion with magnitude: at all radii, the dispersions are highest for the fainter stars and decrease with increasing brightness. The changes in dispersion with magnitude are much less significant than the changes with radius, however they are still present. So the magnitude cuts that we made in \autoref{sect:select_bright} have mitigated the effect of magnitude on dispersion but have not removed it completely.

For our current analysis, this is sufficient. We wish to look at general kinematic properties of the clusters and we are able to do so with the cuts that we have made. However, any modelling of the underlying physics must account for both the incompleteness of the sample that we noted when considering the number counts in each bin and the effects of magnitude on the kinematics that we have noted here.

Finally, in the right panel of \autoref{fig:kin2d_rmag}, the pixels are coloured according to the velocity anisotropy of the stars in the bin to which the pixel belongs from red (high) to blue (low), where the colour bar is symmetric so that green represents isotropy. As for the 2D spatial study, we do not discern any trends in anisotropy with either magnitude (mass) or radius. Again, we believe that the change from isotropy to mild radial anisotropy that we noticed in the 1D profile is lost in the pixel-to-pixel noise.

In future papers, where we will not limit ourselves to only bright stars of similar mass, we will investigate the effects of both mass and position on the kinematics in detail. The cursory analysis we have performed here demonstrates the potential of these beautiful datasets.


\section{Discussion}
\label{sect:discussion}

We wish to compare our dispersion estimates against previous studies. A comprehensive comparison, from which we will derive dynamical distance estimates, will be the subject of a future paper. Here we make only a basic comparison. We also use the kinematic properties we have measured here to look for trends and correlations with other cluster properties, like those we listed earlier in \autoref{table:basics}.

\subsection{Comparison with literature dispersion estimates}
\label{sect:dispn_lit}

We begin by comparing the central dispersions estimated by our polynomial fits with the central dispersion estimates provided in the \citetalias{harris1996} catalogue\footnote{NGC\,5927 is not included in this comparison as no central dispersion estimate is available in \citetalias{harris1996}.}. In order to compare our dispersions in \masyr\ to the \citetalias{harris1996} estimates in \kms, we use the distances from \citetalias{harris1996} to make the appropriate conversions. Both the distances and dispersions are listed in our summary of cluster properties in \autoref{table:basics}.

The dispersion estimates from \citetalias{harris1996} come from different sources and, in some cases, extrapolation was required in order to estimate central dispersions or the dispersion quoted is that inside some central aperture. Furthermore, although we have limited our study here to only bright stars, it is still likely that our datasets probe subtly different stellar populations. Finally, we consider that the distances we have used to convert our proper motions into transverse velocities may also be subject to error. As such, we expect our results to be similar to the \citetalias{harris1996} values, but not identical.

\begin{figure}
    \centering
    \includegraphics[width=\linewidth]{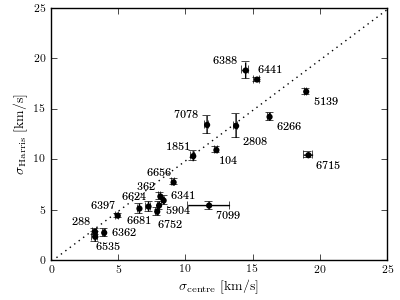}
    \includegraphics[width=\linewidth]{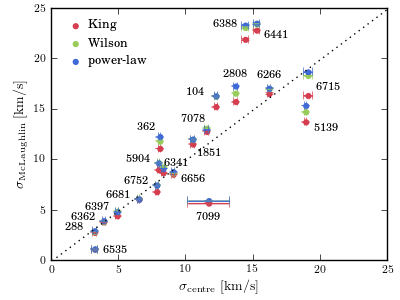}
    \caption{Comparisons of our central dispersions with literature estimates. Top: central dispersion from \citetalias{harris1996} versus our central dispersion. The line for which the estimates are equal is marked as a dotted line. Bottom: central dispersion from \citetalias{mclaughlin2005} versus our dispersion.}
    \label{fig:litcomp}
\end{figure}

We show the results of our comparison in the top panel of \autoref{fig:litcomp}; to guide the eye, a dotted line highlights where the two estimates are equal. In general, our central dispersions are in good agreement with the \citetalias{harris1996} estimates as most points fall along the 1:1 correspondence line with little scatter, though we note that we do tend to overestimate the central dispersions more than we underestimate. We believe this is due to the difference in the estimation process mentioned earlier. Our estimates are taken at $R = 0$, whereas the \citetalias{harris1996} estimates were often extrapolated from dispersions further out or were taken from papers that calculated the dispersion within some central aperture; as dispersion decreases with radius, these methods will tend to underestimate the central dispersion estimates. Further, although we imposed a brightness limit on each of the clusters in order to mitigate the effect of stellar mass on velocity dispersions, we still expect that stars 1 magnitude below the MSTO will have slightly higher dispersions than stars on the RGB; this may also serve to slightly increase our dispersions.

There are two obvious outliers for which we estimate markedly higher dispersions: NGC\,6715 and NGC\,7099. The former is by far the most distant cluster in our sample (at 26.5~kpc) and sits right at the centre of the Sagittarius dwarf spheroidal galaxy. The centre is, therefore, very crowded and cleaning this dataset of unreliable stars was particularly challenging. Including blended or poorly measured stars will tend to increase the dispersion estimates, so it is possible that our cleaning algorithms have not been sufficient and that our dispersion estimate is high as a result. However, although challenging, we are confident that this dataset has been cleaned satisfactorily. So we believe that the discrepancy here is due to the caveats mentioned previously.

NGC\,7099 has the smallest dataset of all 22 clusters in our study; we are left with only 123 stars after cleaning, from which we are able to make just four bins in our 1D analysis. The polynomial fit to this cluster is poorly constrained and with large scatter, as indicated by the large error bar in the figure. However, we do not believe that this is the source of the discrepancy. A central dispersion of 5.5~\kms for an object at a distance of 8.1~kpc \citepalias[both from][]{harris1996}, corresponds to 0.14~\masyr. This is markedly lower than the dispersion profile that we measure for NGC\,7099 (see \autoref{fig:kin6}). Recall that, for this cluster, it was not possible to calculate local corrections; we believe that local inhomogeneities have added extra scatter to the proper-motion measurements and artificially increased our dispersion estimates.

As a second test of our results, we compare our central dispersion predictions with those from \citetalias{mclaughlin2005}\footnote{NGC\,5927 and NGC\,6624 are not included in this comparison as they were not part of the \citetalias{mclaughlin2005} study.}. They fitted single-mass models to cluster surface-brightness profiles using mass-to-light ratios from population-synthesis models, and ages and metallicities from CMD studies, then used the fits to derive structural parameters and predict kinematic properties. They used three classes models -- King models, Wilson models and power-law models -- and we consider all three predictions here. It is worth noting that their results pertain to `average' and not `bright' stars, so we expect that their dispersions may be higher, on average.

We show the results of our comparisons in the bottom panel of \autoref{fig:litcomp}; the King-model predictions are shown as red points, the Wilson-model predictions as green points and the power-law-model predictions as blue points. The dotted line highlights where our central dispersion estimates and the \citetalias{mclaughlin2005} model predictions are equal. As before, we use the distance estimates from \citetalias{harris1996} to convert our dispersions estimate from \masyr\ to \kms.  Once again, the dispersions are generally in good agreement, as most points fall along the 1:1 correspondence line with little scatter. However, we note that now we tend to underestimate the central dispersions more than we overestimate; as noted above, this is not unexpected given the details of the \citetalias{mclaughlin2005} modelling.

We consider briefly the two outliers from the \citetalias{harris1996} comparison. Unlike before, our estimate for NGC\,6715 is in very good agreement with the \citetalias{mclaughlin2005} predictions; this lends further weight to our assertion that our data cleaning and subsequent analysis were proficient and that the previous discrepancy is due to differences in the details of the central dispersion estimations. NGC\,7099, on the other hand, remains an outlier in this comparison; as previously discussed, we believe this is due to our inability to perform local corrections for this cluster.

There are two additional clusters that are outliers in this comparison: NGC\,6388 and NGC\,6441. These are two of the most crowded clusters in our sample; as previously discussed, crowding affects fainter (less massive) stars more significantly than brighter (more massive) stars. The crowding is so high in these clusters that no faint stars survive our quality selection cuts in the central regions. Although we made a magnitude cut to lessen the effect of stellar mass on the velocity dispersions we calculate, we are still left with a small range of masses. Due to energy equipartition, the brighter stars will have a lower velocity dispersion than the fainter stars; it follows then that, if we have removed all the faint stars from the centre, we will tend to underpredict the velocity dispersions there.

\subsection{Dispersion and concentration}
\label{sect:dispn_conc}

The central concentration of a cluster is typically described using the King concentration parameter $c = \log \left( R_\mathrm{tidal} / \Rcore \right)$. We expect that a cluster with a dense core (i.e. a high $c$ value) will have a dispersion that falls off more quickly with radius. We test this prediction using a rather crude approximation of the dispersion slope: the dispersion at the core radius $\score = \sigma (\Rcore)$ as a fraction of the dispersion at the half-light radius $\shalf = \sigma (\Rhalf)$.

\begin{figure}
    \centering
    \includegraphics[width=\linewidth]{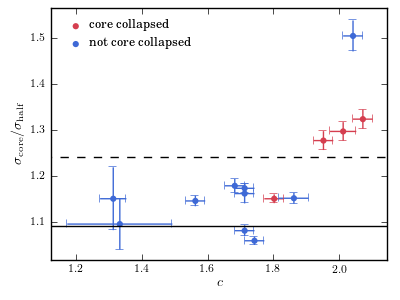}
    \caption{The ratio of the dispersion measured at the core radius $\score$ to that measured at the half-light radius $\shalf$ as a function of cluster concentration, as listed in \citetalias{mclaughlin2005}. The red points are clusters believed to be core collapsed and the blue points are clusters not believed to be core collapsed. $\score/\shalf$ is a crude proxy for the slope of the dispersion profile. There is a clear correlation such that more concentrated clusters have steeper dispersion profiles. For comparison, the solid line shows the value of $\score/\shalf$ expected for an isotropic Plummer model and the dotted line that for a fully radial model \citep{dejonghe1987}.}
    \label{fig:corehalf}
\end{figure}

\autoref{fig:corehalf} shows core-half dispersion ratio $\score/\shalf$ and concentration parameter $c$ (see \autoref{table:basics}) for our clusters. Clusters identified as core collapsed in \citetalias{mclaughlin2005} are shown in red \citep[we also include NGC\,362, which was identified as core collapsed by][]{dalessandro2013} and those not identified as core collapsed are shown in blue. As expected, we see a clear trend of dispersion ratio with concentration, indicating that the most centrally-concentrated clusters do indeed have steeper dispersion slopes than the looser clusters. The point with very high $\score/\shalf \sim 1.5$ is NGC\,6715; as previously discussed, this cluster sits right at the centre of the Sagittarius dwarf spheroidal, which may explain both its high central concentration and the sharp fall of its dispersion profile, even though it is not believed to be core collapsed.

By way of comparison, we show the value of $\score/\shalf$ expected for an isotropic Plummer model as a solid line. Radially-anisotropic Plummer models \citep{dejonghe1987} would have higher $\score/\shalf$ values; as an upper limit, we show the value expected for fully radial models as a dashed line. We see that the clusters with low or moderate central concentration are generally described well by this family of models, whereas the clusters with high central concentration are not well described as they have higher values of $\score/\shalf$ even than models with purely radial orbits.

\subsection{Anisotropy and ellipticity}
\label{sect:aniso_ellip}

In order to estimate anisotropy, we require two orthogonal components of velocity, so line-of-sight velocity studies alone cannot give us this information; we need proper motions. As so few globular clusters have been studied with proper motions before, it is not possible to make a comparison of our anisotropy measurements with previous estimates. However, we can use the anisotropies we have calculated and look for correlations with other cluster statistics. In what follows, we use only the clusters with over 1000 stars remaining in their cleaned samples, thus excluding NGC\,6535 and NGC\,7099, as we found that the noise overwhelmed the signal for the anisotropy profiles for these very small datasets.

First, we consider cluster ellipticity $\epsilon$ and minor-axis/major-axis anisotropy $\sminor / \smajor$. Ellipticity describes the difference in the major- and minor-axis lengths for the cluster, where $\epsilon=0$ indicates a perfectly round cluster and increasing $\epsilon$ indicates increasingly flattened clusters. The minor-major anisotropy reflects the difference in the velocity dispersions along the major and minor axes, where $\sminor / \smajor = 1$ indicates isotropy, $\sminor / \smajor < 1$ indicates a preference for motion along the major axis and $\sminor / \smajor > 1$ indicates a preference for motion along the minor axis. 

To determine the major-axis and minor-axis directions, we use position angles from \citet{white1987}. We then calculate the one-dimensional major-axis dispersion $\smajor$ and minor-axis dispersion $\sminor$ profiles in the same bins that were used for the analysis of the one-dimensional kinematics in \autoref{sect:kin1d}. Then the minor-major anisotropy is simply $\sminor / \smajor$. In order to have a single, representative value (with uncertainty) against which to compare the ellipticities, we take the weighted mean (and error on the weighted mean) of the anisotropies from the binned profile, using the inverse square uncertainties as weights. The anisotropy $\sminor / \smajor$ thus inferred represents an average over the entire radial range for which we have data.

We now wish to fit a straight line to the anisotropies as a function of ellipticity to identify any correlations and assess their significance. To begin, we do this using a simple least-squares method -- this assumes that the data have been drawn from the model (in this case a straight line) with the given error bars. If the uncertainties have been well estimated, then the $\chi^2$ calculated using the best-fitting model should be $\sim N$, where $N=19$ is the number of clusters for which we are able to carry out this analysis. However, we find that the best fit has $\chi^2 \gg N$. This implies that there are other sources of scatter in addition to random errors. To correct for this, we increase all error bars by a factor $\sqrt{\chi^2_\mathrm{best}/N}$ before re-fitting a straight line to the data points. Increasing the size of the error bars does not change the best-fit line, only the uncertainties on the fitted parameters.

\begin{figure}
    \centering
    \includegraphics[width=\linewidth]{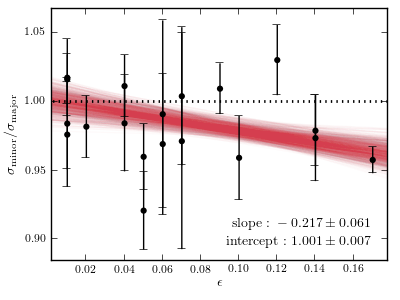}
    \caption{Minor-axis/major-axis anisotropy $\sminor/\smajor$ versus ellipticity $\epsilon$. The points represent our clusters and the dotted line indicates isotropy where $\sminor/\smajor = 1$. The red lines are draws from a straight-line fit to the data and highlights a mild correlation. Round clusters ($\epsilon = 0$) are isotropic and flattened clusters show mild major-axis anisotropy $\sminor/\smajor < 1$, with the degree of anisotropy increasing with ellipticity. We note that the fit was not forced through (0,1).}
    \label{fig:aniso_ellip}
\end{figure}

\autoref{fig:aniso_ellip} shows minor-major anisotropy and cluster ellipticity\footnote{NGC\,288 is not included in this comparison as it has neither an ellipticity listed in \citetalias{harris1996} nor a position angle listed in \citet{white1987}.}. The dotted line at $\sminor/\smajor = 1$ represents isotropy. The red lines show draws from the straight-line fit; we also give the values of the fit parameters. Note that both slope and intercept are left completely free in the fit, we do not force any special behaviour. We see a mild correlation between ellipticity and minor-major anisotropy. Round clusters ($\epsilon = 0$) are isotropic, while flattened clusters show a small degree of major-axis anisotropy, that is, their velocity dispersions tend to be larger along the major-axis than along the minor-axis. So as clusters are elongated along the major axis, it seems that so too is their velocity distribution, although a full 3D velocity study, including rotation, would be necessary to characterise the shape and direction of the velocity ellipsoid.

\subsection{Anisotropy and relaxation time}
\label{sect:aniso_relax}

Violent relaxation during the early stages of cluster formation typically leads to systems that are isotropic at their centres and become radially anisotropic with increasing distance from the cluster centre \citep{lyndenbell1967}. This behaviour has been clearly demonstrated by N-body simulations of isolated galaxies and clusters \citep[e.g.][]{vanalbada1982, trenti2005} and also, recently, in clusters evolving in an external tidal field \citep{vesperini2014}. Isolated clusters simply become increasingly anisotropic with radius, however \citet{vesperini2014} noted that tidal effects cause the anisotropy profiles to turn over in the outer regions and become isotropic or even mildly tangential.

Violent relaxation is a collisionless process driven by fluctuations in the cluster potential during collapse. Thereafter, once the cluster has reached a steady state, the system undergoes collisional relaxation: over time, as stars experience two-body interactions that affect small changes to their orbits, their motions become more random, i.e. they move towards isotropy. In the inner regions, where the relaxation times are short, we would expect the velocity distributions to be approximately isotropic. However, in the intermediate and outer regions, where the relaxation times are longer, it is possible that clusters are not yet fully relaxed and so we would expect to still observe some radial anisotropy. Further, we expect that the degree of radial anisotropy observed, if any, will depend on the relaxation time.

We observe this general behaviour in the 1D anisotropy profiles in Figures~\ref{fig:kin1}-\ref{fig:kin6}: most clusters appear to be nearly isotropic (with $ \st/\sr = 1$) at their centres and show some radial anisotropy ($\st/\sr < 1$) further out. Our catalogues lack the spatial coverage to fully probe the behaviour of the anisotropy profiles in the outermost regions, but we are well able to study the inner and intermediate regions.

\begin{figure}
    \centering
    \includegraphics[width=\linewidth]{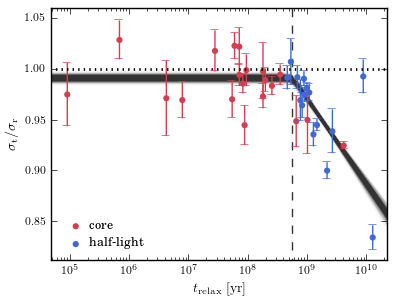}
    \caption{Tangential/radial anisotropy $\st/\sr$ versus relaxation time $\trelax$. The dashed line represents isotropy where $\st/\sr=1$. The red points show core relaxation times and anisotropies estimated at core radii; the blue points show half-mass relaxation times and anisotropies estimated at half-light radii. Regions with short relaxation times have had time to become fully relaxed and have isotropic velocity distributions, this is the case for most cluster cores. Regions with longer relaxations times are not yet fully relaxed and so show mild radial anisotropy, with the degree of anisotropy increasing with relaxation time. The black lines are draws from a functional fit to the points, as described in the text. The dotted line marks a critical relaxation time beyond which velocity distributions have not yet had time to become fully relaxed. The blue outlier in the top right of the figure is NGC\,6715 (M\,54), the most distant in our sample.}
    \label{fig:trelax}
\end{figure}

So can we relate the anisotropies directly to the relaxation times of the clusters? \autoref{fig:trelax} shows $\st/\sr$ anisotropy as a function of the relaxation time (see \autoref{table:basics}); once again, we exclude NGC\,6535 and NGC\,7099 as those have fewer than 1000 stars remaining after cleaning. The dotted line indicates isotropy. The red points show core relaxation times and anisotropies estimated at the core radius; the blue points show half-mass relaxation times and anisotropies estimated at the half-light radius. As expected, we see that short relaxation times result in isotropic distributions, whereas longer relaxation times lead to distributions that show mild radial anisotropy. Most clusters appear to have relaxed cores, but very few are relaxed out to their half-light radius.

We assume that for relaxation times shorter than some limiting time $\tbreak$, the velocity distributions are isotropic, and that for relaxation times longer than $\tbreak$, there is some radial anisotropy, which increases in strength for longer relaxation times. So, to the combined sample of core (red) and half-light (blue) estimates, we fit a function
\begin{equation}
    f \left( t \right) = \left\{ \begin{array}{ll}
        C & \mbox{$t \le \tbreak$} \\
        C + S \left( \log{t} - \log{\tbreak} \right) & \mbox{$t > \tbreak$}
       \end{array} \right.
\end{equation}
for constant $C$ and slope $S$. We expect that $C$ will be very close to 1, but we do leave it free in the fit. In \autoref{fig:trelax}, the solid lines show draws from the final fit distribution. We find $C \sim 0.992$, so very close to isotropy, as predicted; in this relaxed regime, the cluster-to-cluster rms is just 0.021, and so the error on $C$ is $\Delta C \approx 0.021/\sqrt{N} = 0.005$. The dashed line marks the best-fit value of $\tbreak \sim 0.55$~Gyr; this implies that only regions of clusters with shorter relaxation times have had time to fully relax. This may serve as a useful indicator of relaxation in other, less-studied clusters.

\subsection{Kinematic Centres}
\label{sect:kincentres}

In \autoref{sect:kin2d}, we extracted two-dimensional velocity dispersion maps as a function of spatial coordinates. We now use those maps to make a rough estimation of the kinematic centres. The estimates we make here are crude, and are simply designed to serve as a test of our methodology.

Recall, to make the original dispersion maps, we binned the stars into a grid of $30 \times 30$ pixels, grouped the pixels into Voronoi bins and then calculated the velocity dispersion in each bin. Then each pixel was assigned a dispersion (and uncertainty) equal to the dispersion (and uncertainty) of all the stars in the bin to which that pixel belonged. To estimate the kinematic centre, we begin by re-pixelating the dispersion map into a $150 \times 150$ pixel grid. We assign a dispersion to each new pixel by drawing at random from the dispersion distribution of the nearest original pixel. As well as increasing the resolution of the pixel grid, this also fills in empty pixels from the original grid where there were no stars; this extrapolation is not ideal but is necessary to avoid boundary effects.

\begin{figure*}
    \centering
    \includegraphics[width=0.33\linewidth]{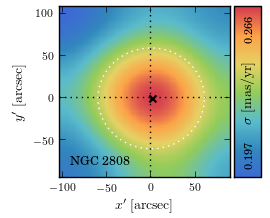}
    \includegraphics[width=0.33\linewidth]{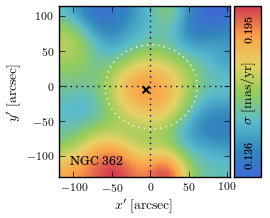}
    \includegraphics[width=0.33\linewidth]{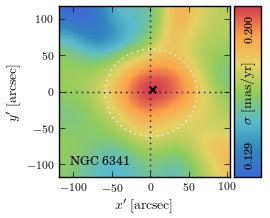}
    \caption{Smoothed 2D velocity dispersion maps for NGC\,2808, NGC\,362 and NGC\,6341 as a function of spatial coordinates. Each pixel is coloured according to the gaussian-smoothed dispersion from red (high) to blue (low), as indicated by the colour bars. The black dotted lines mark $x'=0$ and $y'=0$; their intersection marks the photometric cluster centre adopted for this study \citepalias[see Table 1 of][]{bellini2014}. The white dotted circles mark the central arcminute. The kinematic centre estimates are shown as black crosses and represent the peak of the smoothed dispersion profile inside of the central arcminute.}
    \label{fig:kin2d_smooth}
\end{figure*}

We then apply a Gaussian-smoothing filter of width $N$ pixels to the new pixel grid. After some trial-and-error, we found that some clusters worked best with $N=12.5$ and others worked best with $N=25$, depending on the properties of the cluster. The smoothed dispersion map for NGC\,2808 is shown in the left panel of \autoref{fig:kin2d_smooth} as a function of spatial coordinates on the plane of the sky. The colour bar shows the magnitude of the smoothed dispersion from high (red) to low (blue). The black dotted lines highlight the photometric centre at $(0,0)$ \citepalias[see Table 1 of][]{bellini2014}.

We estimate the kinematic centre to be the pixel with the highest dispersion in the smoothed grid that lies within the central arcminute. In \autoref{fig:kin2d_smooth}, the cross marks the position of the estimated kinematic centre and the white dotted line delineates the central arcminute. We impose the radius limit because, for some clusters, we see high dispersions at the edges of our fields, which we attribute to edge effects. We also note that the outer pixels are less reliable than the inner pixels as coverage in the original grid was more sparse there, and in some cases the nearest original pixel used to assign a dispersion to a new pixel was actually not very nearby at all; again, the radius limit helps to mitigate this effect. There were four clusters for which we were not able to estimate a kinematic centre due to the poor quality of the smoothed dispersion map: NGC\,5904, NGC\,6362, NGC\,6397 and NGC\,6535.

To estimate the uncertainty on our centre estimates, we fit a half-gaussian to the smoothed dispersion as a function of distance from the estimated centre for all pixels within the central arcminute of the photometric centre. If the width of the fitted half-gaussian is $W$ then we estimate the uncertainty on the centre to be $W/\sqrt{N(R<W)}$, where $N(R<W)$ is the number of stars in our dataset within a radius of $W$. The uncertainty is shown in \autoref{fig:kin2d_smooth} but is too small to be distinguished on the scale of the plot. For NGC\,2808, we estimate the centre at $(x',y') = (1.9 \pm 0.9,-0.7 \pm 0.9)$~arcsec.

\begin{figure}
    \centering
    \includegraphics[width=\linewidth]{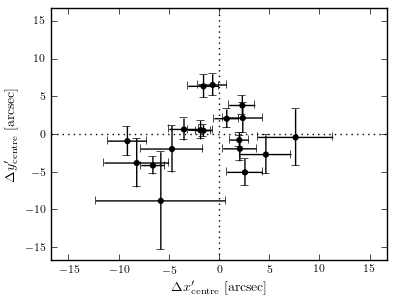}
    \caption{Kinematic centres estimates for 18 of our clusters. The centres were estimated by gaussian-smoothing the dispersion maps and searching for the peak of the smoothed map within the central arcminute. The dotted lines mark $\left( x',y' \right) = \left( 0,0 \right)$, the position of the photometric centre for each cluster.}
    \label{fig:kin2d_centres}
\end{figure}

NGC\,2808 is one of our cleanest dispersion maps so, in the middle and right panels of \autoref{fig:kin2d_smooth}, we show the smoothed maps for NGC\,362 and NGC\,6341. The former is an example where edge effects artificially increased the dispersion towards the bottom of the field. The latter is an example where the filling in of empty pixels to avoid boundary effects can cause a smearing of the dispersion map in the outer regions. Both highlight the need for the radius limit on the centre search.

\autoref{fig:kin2d_centres} shows the offsets of the centre estimates for our clusters, with dotted lines highlighting the adopted photometric centres at $x'=0$ and $y'=0$. The unweighted RMS of the centre offsets in the $x$ and $y$ directions are $4.60"$ and $3.79"$, respectively. The unweighted mean and error-in-the-mean of the centre offsets for the sample as a whole are $(\overline{\Delta x'},\overline{\Delta y'}) = (1.15" \pm 1.08", -0.35" \pm 0.89")$. Therefore, at 1$\sigma$ confidence in two dimensions, the photometric and kinematic centres of globular clusters agree on average to within $\sim 1"$.


\section{Conclusions}
\label{sect:conclusions}

We have performed a kinematical analysis of 22 Milky Way globular clusters using the \textit{HST} proper-motion catalogues described in \citetalias{bellini2014}. Approximately half of our cluster sample have been previously studied using line-of-sight velocity data, and only a handful have been previously studied using proper motions. So for most clusters, this is the first proper-motion study undertaken and, for many, this is the first kinematical study of any kind.

We began with careful cleaning of our datasets. The quality of any kinematical analysis depends on the quality of the data used for the analysis: underestimated uncertainties, velocity uncertainties larger than the local signal, and the presence of contaminants can all lead to biases in the estimated velocity moments. We also selected only bright stars -- with a small range of masses -- in order to limit ourselves to the study of changes in kinematics with position. Changes in kinematics with mass and with both position and mass will be the subject of future papers.

Analogous to typical line-of-sight velocity studies, we determined binned velocity-dispersion profiles as a function of radius for all clusters. To these profiles, we fitted a polynomial and then used the fit to estimate the dispersions at the cluster centre, the core radius and the half-light radius. We compared the central dispersions to estimates in \citetalias{harris1996} and predictions from \citetalias{mclaughlin2005} and found our results to be in very good agreement. We then used the ratio of the core and half-light dispersion estimates as a proxy for dispersion slope and compared the slopes to cluster central concentrations. We found a mild correlation between dispersion slope and concentration, indicating that clusters with high central concentration have dispersion profiles that fall off more quickly with radius.

We went on to determine binned tangential/radial velocity anisotropy profiles as a function of radius for all clusters. Such an analysis requires two components of velocity information and so is not possible for studies using only line-of-sight velocities. This is the first time it has been possible to determine anisotropies for most of the clusters in our sample, which will be crucial for breaking mass-anisotropy degeneracy in future dynamical modelling studies. We found that most of the clusters studied here are isotropic in the centre and become mildly radially anisotropic in their outer regions. This is understandable if we consider that stellar orbits in clusters are preferentially radial at formation and move towards isotropic as the stars undergo two-body interactions and the cluster relaxes. Given sufficient time to become fully relaxed, the stellar velocity distributions in the inner regions will become fully isotropic. In the outer regions of clusters, where tidal effects become important, stars will move away from isotropy \citep{vesperini2014}; however, our catalogues are focused in the central regions of the clusters and lack the coverage to probe such effects.

Further, we examined our anisotropy estimates as a function of relaxation time and found that only regions of clusters with $\trelax < 0.55$~Gyr -- primarily the cluster cores -- have had time to relax fully and reach isotropy. For relaxed regions, we find a mean anisotropy $\st/\sr \sim 0.992$, with a cluster-to-cluster rms scatter of 0.021. We also investigated minor-axis/major-axis velocity anisotropy as a function of cluster ellipticity and found a clear correlation whereby round clusters are isotropic and flattened clusters have flattened velocity ellipsoids on the plane of the sky.

Finally, we determined two-dimensional spatially-binned dispersion maps for all clusters. Such an analysis is possible in theory for line-of-sight studies, but is often impossible in practice due to the limited sample sizes that naturally arise due to the small number of spectra that can be taken at any one time, even with the most advanced instruments. So, once again, this analysis is a first for most of the clusters in our sample. For NGC\,6535 and NGC\,7099, our datasets are too small to determine meaningful information; for the remaining 20 clusters, we were able to obtain beautiful dispersion maps that clearly show peaks of high velocity dispersion at their centres that fall of with radius. By Gaussian smoothing the dispersion maps, we were also able to identify kinematic centres for all but four of the clusters.

As we have already discussed, much of the kinematical analysis presented here has been the first of its kind for many of the clusters in our sample. And yet, we have still only scratched the surface of what is possible with the data we currently have and that we anticipate over the coming years \citep[see discussion in][]{bellini2014}. Future papers will perform detailed dynamical modelling to connect the observations to the underlying physics; this will provide greater insight into the dynamical structures of the clusters and truly unlock the potential of these remarkable datasets.


\section*{Acknowledgements}

We would like to thank Paolo Bianchini for interesting discussions on the topic of data quality and cleaning. We are also grateful to Ivan King, Julio Chanam\'{e}, Rupali Chandar, Adrienne Cool, Francesco Ferraro, Holland Ford, Davide Massari and Giampaolo Piotto for collaboration on other aspects of this project. Finally, we would like to thank the anonymous referee for their careful reading of the manuscript and for their useful comments.

This research made use of Astropy\footnote{\url{http://www.astropy.org}}, a community-developed core Python package for Astronomy \citep{astropy2013}.


\bibliographystyle{apj}

\bibliography{refs}


\clearpage
\appendix

\section{One-dimensional kinematic profiles}
\label{sect:kin1d_profiles}

\autoref{table:kin1d_profiles} gives the binned one-dimensional proper-motion dispersion and anisotropy profiles derived in \autoref{sect:kin1d} and shown in Figures~\ref{fig:kin1d},\ref{fig:kin1}-\ref{fig:kin6}. The table contains results for all clusters together. The columns list: the number of stars in the bin; the mean and standard deviation of the radii of the stars to give an indication of bin position and width; the velocity dispersion, with uncertainty, of the stars in the bin; and the velocity anisotropy, with uncertainty, of the stars in the bin.

\begin{table}[h]
    \centering
    \caption{One-dimensional kinematic profiles.}
    \label{table:kin1d_profiles}
    \begin{tabular}{ccccccccc}
        \hline
        \hline
        Cluster & Bin & $N$ & $R$ & $\Delta R$ & $\sigma$ & $\Delta \sigma$ & $\st/\sr$ & $\Delta \st/\sr$ \\
        & & & \multicolumn{2}{c}{(arcsec)} & \multicolumn{2}{c}{(\masyr)} & \multicolumn{2}{c}{} \\
        (1) & (2) & (3) & (4) & (5) & (6) & (7) & (8) & (9) \\
        \hline
        NGC\,104 
        &   1 &   25 &   7.945 &   3.176 & 0.597 & 0.060 & 1.458 & 0.294 \\
        &   2 &   44 &  13.464 &   1.323 & 0.593 & 0.047 & 1.154 & 0.165 \\
        &   3 &   43 &  16.841 &   0.790 & 0.579 & 0.046 & 1.386 & 0.202 \\
        &   4 &   44 &  19.019 &   0.585 & 0.568 & 0.043 & 0.774 & 0.116 \\
        &   5 &  793 &  25.746 &   2.627 & 0.559 & 0.009 & 1.035 & 0.037 \\
        &   6 &  792 &  32.928 &   1.764 & 0.579 & 0.011 & 0.984 & 0.033 \\
        &   7 &  792 &  38.188 &   1.397 & 0.565 & 0.011 & 0.957 & 0.037 \\
        &   8 &  792 &  42.988 &   1.376 & 0.553 & 0.010 & 1.004 & 0.037 \\
        &   9 &  792 &  47.423 &   1.201 & 0.562 & 0.009 & 0.994 & 0.036 \\
        &  10 &  792 &  51.638 &   1.242 & 0.537 & 0.009 & 0.956 & 0.036 \\
        \hline
    \end{tabular}
    
    \qquad
    
    \textbf{Notes.} Columns: (1) cluster ID; (2) bin number; (3) number of stars; (4-5) mean radius and uncertainty; (6-7) combined radial and tangential velocity dispersion and uncertainty; (8-9) tangential/radial anisotropy and uncertainty.
    
    This table is published in its entirety in a machine-readable form in the online journal. A portion is shown here for guidance regarding its form and content.
\end{table}


\end{document}